\documentclass[useAMS,usenatbib]{mn2e}
\usepackage{amsmath}    % need for subequations
\usepackage{amssymb}
\usepackage{journals}
\usepackage{graphicx}   % need for figures
\usepackage{verbatim}   % useful for program listings
\usepackage{color}      % use if color is used in text
\usepackage{ulem}		%be able to use nicer editing option
\title[Exploring plasma evolution during Sgr A* flares]{Exploring plasma evolution during Sagittarius A* flares}
\author[Salom\'e Dibi et al.]{S. Dibi$^{1}$\thanks{Salom\'e Dibi's e-mail:
s.dibi@astro.ru.nl},  S. Markoff$^{1}$, R. Belmont$^{2,3}$,  J. Malzac$^{2,3}$, N. M. Barri\`ere$^{4}$, and J. A. Tomsick$^{4}$ \\\\
$^{1}$Astronomical Institute ``Anton Pannekoek", University of Amsterdam, Postbus 94249,
1090 GE Amsterdam, The Netherlands\\
$^{2}$Universit\'e de Toulouse; UPS-OMP; IRAP; Toulouse, France \\
$^{3}$CNRS; IRAP; 9 Av. colonel Roche, BP 44346, F-31028 Toulouse cedex 4, France \\
$^{4}$Space Sciences Laboratory, 7 Gauss Way, University of California, Berkeley, CA 94720, USA}

\begin{document}

\date{Accepted in MNRAS on March 25, 2014. MNRAS 441, 1005–1016 (2014)} 

\pagerange{\pageref{firstpage}--\pageref{lastpage}} \pubyear{2014}

\maketitle

\label{firstpage}

\begin{abstract}

We present a new way of describing the flares from Sgr A* with a self-consistent calculation of the particle distribution. All relevant radiative processes are taken into account in the evolution of the electron distribution and resulting spectrum. We present spectral modelling for new X-ray flares observed by NuSTAR, together with older observations in different wavelengths, and discuss the changes in plasma parameters to produce a flare.

We show that under certain conditions, the real particle distribution can differ significantly from standard distributions assumed in most studies.

We conclude that the flares are likely generated by magnetized plasma consistent with our understanding of the accretion flow.  Including non-thermal acceleration, injection, escape, and cooling losses produces a spectrum with a break between the infrared and the X-ray, allowing a better simultaneous description of the different wavelengths. We favour the non-thermal synchrotron interpretation, assuming the infrared flare spectrum used is representative.

We also consider the effects on Sgr A*’s quiescent spectrum in the case of a density increase due to the G2 encounter with Sgr A*.

\end{abstract}

\begin{keywords}
Galaxy: center -- Galaxy: nucleus --  accretion discs -- black hole physics -- MHD  -- radiation mechanisms: thermal -- relativistic processes -- methods: numerical  -- galaxies: jets.
\end{keywords}

\section{Introduction}

 \subsection{Sagittarius A*}

	Sagittarius A* (Sgr A*) is the name given to the bright radio source of our Galactic Center. It was discovered in 1974 by \citet{balick74} using the Green Bank 35 km radio link interferometer of the National Radio Astronomy Observatory.  Stellar motion around the non-thermal radio source shows that Sgr A* is highly compact (smaller than 0.01 pc i.e. $3\times 10^{11}$km) and that the stars orbit around a point mass of $4.3\pm0.5\times10^6M_\odot$ \citep{eisenhauer05, melia07}. The stellar orbits provide the strongest evidence yet for a supermassive black hole located in the center of our Galaxy at a distance of $8.3\pm0.35~{\rm kpc}$ \citep{reid93, schodel02, ghez08, gillessen09}. Super-massive black holes (SMBHs) of millions to billions of solar masses are believed to exist in the centre of most galaxies. Sgr A*, in our own galaxy, is the closest and best studied SMBH, making it the perfect source to test our understanding of galactic nuclei systems in general. But among all galactic nuclei that we have observed so far, Sgr A* has the peculiarity of being very faint in all wavelengths. Even though it may have been more active in the past \citep{revnivtsev04, zubovas12, ponti12}, today Sgr A* is one of the most under-luminous SMBHs we know, it is very faint with  $L_{\rm bol}\simeq 10^{-9} L_{\rm Edd}$ \citep{narayan98} and accreting at a very low rate. The accretion rate has been constrained by polarisation measurements, using Faraday rotation \citep{aitken00, bower03, marrone07} and is estimated to be in the range $2\times10^{-9}<\dot{M}<2\times10^{-7} M_{\odot}~{\rm yr^{-1}}$. Theoretical work suggest that Sgr A* is most likely accreting at the lower range of this interval \citep{moscibrodzka09, drappeau13}. \citet{dibi12} have shown that for accretion rate above or equal to $1 \times 10^{-8}$ solar masses per year, the cooling losses become important in the modelling of the accretion flow and the resulting spectrum. This result means that Sgr A* is the only black hole source known where cooling could still be treated separately as a first approximation. Along the same lines, \citet{yuan04} had shown that flare events, as these observed in the Galactic Centre, could not be detected if the accretion rate increases by a factor 10 from its actual value, because synchrotron self Compton emission from thermal electrons would increases substantially. This would explain why Sgr A* is the only source known to exhibit flaring activity. Also, \citealp{yusefzadeh09} are reporting observational evidence for IR flaring activity inversely proportional to the flux density.
	
 \subsection{Multiwavelength observations}	

The faint emission from Sgr A* has been observed in different wavelengths giving us a broad band spectrum of this object from the radio to the X-ray (see reviews by \citealt{ melia01, genzel10, morris12}, and references therein). From a few GHz up to 100 GHz, the radio spectrum extends as a rough power law $F_{\nu } \propto \nu ^{\alpha }$ with 0.25 $\leq \alpha \leq$ 0.33. Above 100 GHz there is evidence for a millimeter/sub-millimeter (sub-mm) excess over the power law, extending almost to 1000 GHz. The nature of this excess was discussed by \citet{serabyn97} and \cite{falcke98} who excluded the possibility of dust emission. The size of photosphere is predicted to be the smallest around this wavelength of 1.3 mm, where the excess is observed. And the black hole horizon, or its shadow  \citep{falcke00,dexter10} could be detected for the very first time in the near future with new, very long base interferometry facilities such as the ``Event Horizon Telescope'' \citep{doeleman08, doeleman09, fish11}.

Important progress has been achieved in the mid-infrared (MIR) to near infrared \citep[NIR; e.g.,][]{genzel03, ghez05, schodel11} and sub-mm domains. But in the optical and in the ultra-violet wavelengths, the Galactic Center is heavily obscured by gas and dust with 30 magnitudes of visual extinction. This obscuring medium becomes partially transparent to the X-rays at energies above 2 keV. Indeed, Sgr A* has a quiescent X-ray luminosity of a few $10^{33} \rm erg \ s^{-1}$ (\citealt{baganoff03}) which is about $10^{11}$ times lower than the Eddington luminosity. 

Sgr A* is quite variable and we observe fast activity (bursts or flares) in the infrared and X-ray band emissions. A few times a day, Sgr A* experiences rapid increases in the NIR flux \citep{hornstein02, genzel03, ghez04, eckart06, eckart08, yusefzadeh08, doddseden11, haubois12}, where brighter flares \citep[$>10$ mJy;][]{doddseden11} are often associated with simultaneous X-ray flares \citep[e.g.][]{baganoff01,goldwurm03,porquet03,belanger05,porquet08, nowak12, neilsen13}.  The typical timescale for such events is few thousand seconds, suggesting a common localized origin of the flares. The radio and sub-mm emissions are more stable, i.e. they show less variability than the X-ray and NIR (see for instance \citealt{marrone08} for the sub-mm flares, and \citealt{yusefzadeh10} for a study on the IR - sub-mm anti-correlation). The flat radio emission is most likely the synchrotron emission originating from an outflow of Sgr A*, the lower the frequency, the further away we are in the outflow. So the radio wavelength, as well as the quiescent X-ray emission are coming from extended regions around Sgr A*, while the sub-mm, MIR, and flaring X-ray emissions originate from a region very close to the SMBH. This second region is the one we are interested in and we explore in this paper.

The MIR and NIR emission has been observed by the VLT and Keck (e.g. \citealt{doddseden11,schodel11,bremer11,witzel12}). The X-ray variability has been observed by XMM-Newton, the Chandra X-ray observatory (e.g. \citealt{baganoff01}) and also by \textit{Swift} \citep{degenaar13}. Many new X-ray flares have been detected recently thanks to the Chandra 2012 Sgr A* X-ray Visionary Project\footnote{http://www.sgra-star.com/} . From this 3-Ms campaign, 39 X-ray flares are reported, lasting from a few hundred seconds to approximately 8 ks, and ranging in 2--10 keV luminosity from $10^{34} \rm ergs \ s^{−1}$ to $2 \times 10^{35} \rm ergs \ s^{−1}$ \citep{nowak12, neilsen13}. The new telescope NuSTAR \citep{harrison13} has released recently new X-ray flares data that we are using in our study (Barri\`ere et al., submitted). Those show that the 3--80 keV emission is compatible with a pure power-law spectrum. 

 \subsection{Flare models}

The fast variability indicates that the origin of the flares is as close as few gravitational radii from the SMBH. However, the nature of the physical processes responsible for the flares is still an open question.  Different mechanisms have been proposed such as events of magnetic reconnection or other acceleration processes (e.g., \citealt{markoff01, yuan03, liu04, liu06}), infall of gas clumps or disruptions of small bodies \citep{cadez06, tagger06, zubovas12b}, adiabatic expansion of hot plasma or hot spot models \citep{yusefzadeh08, broderick06}.
By modelling the physical conditions around Sgr A* and fitting the observational data, we also aim at giving a possible interpretation of the phenomenon. 

Several studies have been devoted to the modelling of SgrA* flares. 
Some models include a precise description of the flow geometry. For instance, the emission from SgrA* was interpreted in the framework of radiatively inefficient accretion flows \citep{yuan03}. In this model, the matter properties (density, temperature, etc.) are computed through hydrodynamical equations including radiative losses. These properties depend on the distance to the black hole and each ring contributes differently to the overall spectrum. 
The outer parts of the accretion flow contribute significantly to the X-ray luminosity in the quiescent state (\citealt{quataert02, baganoff03}) with only about ten per cent of the quiescent X-ray flux coming from the central part we are modelling (\citealt{wang13, neilsen13}).

However, even in models where the geometry is dealt with accuracy, most of the emission originates from the very central parts of the accretion flow, both in the quiescent sub-mm and NIR bands, and in the flaring sub-mm to X-ray bands. 
Moreover, the typical time scale of a flare (few thousand seconds, depending on the flare) is of the order of the orbital period at the inner most stable orbit of SgrA*, pointing again to a flaring region of only a few gravitational radii ($r_G= GM/c^2$).
Therefore, most attempts to model the sub-mm to X-ray spectrum of SgrA* in the quiescent and flaring states (excluding the radio emission) implicitly assume that the emission originates from a single homogeneous, isotropic zone characterized by only few parameters such as the average electron temperature and density, the magnetic field intensity (e.g. \citealt{ doddseden10, liu06}). Here we use the same approach. 

Most models agree on the the fact that the small emitting region is weakly magnetized ($\lesssim$ few hundred Gauss) and faint ($\lesssim 10^8$ particles/cm$^3$) which now appear as standard values (\citealt{dibi12,moscibrodzka09,dexter09}). These values are supported by observations that constrain the accretion rate via Faraday rotation, to a maximum of $\dot{M} \sim 10^{-7}$ solar masses per year. As a simple check, taking this higher accretion rate limit and a ``typical" bulk velocity at one or two gravitational radii of 10$\%$ of the speed of light (as simulated in GRMHD models of Sgr A*), then $\rho_{max} \simeq \dot{M}/(4 \pi R^2 v ) \sim 10^8 \rm particle/cm^3$. 

Whatever the details of the accretion flow and the radiative processes responsible for the emission, the emitted spectrum depends drastically on the particle distribution. For the sake of simplicity, all models so far have assumed pre-determined particle distributions (Maxwellian, power-law, broken power-law, or a combinations of them) which are described by few parameters. The precise shape of the particle distributions depends on the radiative and acceleration processes and can deviate significantly from the assumed ones. The present work aims at dealing more precisely with particle distributions. 
Fitting arbitrary distributions to the data is not possible with current coverage and sensitivity of the instruments. Rather, the shape of the particle distribution can be computed self-consistently with a Boltzmann equation assuming a physics described by a few parameters. Such an approach is common in the modelling of the high energy emission from X-ray binaries and other AGN (e.g. \citealt{mcconnell02, rogers06, belmont08}) but has not been applied to SgrA* yet. In this paper, we present spectra obtained by solving simultaneously an equation for particles and an equation for photons to produce self-consistent particle distributions and spectra. These spectra are compared to broad band data of SgrA* to put constraints on the flare properties. 

This paper is organised as follows: In Section 2 we present the microphysics and numerical method. In Section 3 we present the results and solutions for the quiescent and flaring spectra from Sgr A*. We study the plasma behaviour in two kinds of models, namely in systems where matter is trapped in the emission region, and in systems where matter flows in and out of the emission region. In Section 4 we end with our conclusions and outlook.

\section{Method}

%\subsection{General overview}

The goal of this study is to model the plasma around Sgr A*, and in particular the particle distributions and the resulting spectra. In the following, we will note $\nu$ the frequency of photons, $\gamma$ the particles Lorentz factor , and $p=(\gamma^2-1)^{1/2}$ the particle momentum.

% But we do not assume a given particle distribution as we want to solve self-consistently for this one, taking into account the influence of the different processes on the distribution. Indeed, even if we have an idea of the density range around the SMBH, the shape of the distribution is not constrained by any physical reasons beside trying to reproduce the observed spectra.  We would like to test whether taking arbitrary simple distributions (such as a combination of Maxwellian plus power-law, which is usually done) is a reasonable description of the plasma behaviour. To do so, we need to have as few assumptions as possible on the particle distribution, by solving self-consistently for the particle evolution as this one undergoes different physical processes (radiative cooling, heating, injection, escape, etc.).

We use the {\sc belm} code \citep{belmont08}. This numerical tool solves simultaneously coupled kinetic equations for leptons and photons in a magnetized, uniform, isotropic medium of typical size $R$. In all models presented here, this size is set to $R=2 \ r_G = 1.3 \times 10^{12}$ cm  based on the size derived from the flare time scale variability (where $r_G=GM/c^2$ is the gravitational radius).

%We use the numerical code \emph{BELM}  described in \citet{belmont08} that solves the kinetic equations for the time evolution of the particle population. The code was designed to model the emission properties of high energy plasma sources. The time-dependent kinetic equations are solved for homogeneous and isotropic distributions of photons, electrons and positrons with any initial particle distribution. 

The implemented microphysics includes radiation processes as self-absorbed radiation, Compton scattering, self-absorbed bremsstrahlung radiation, pair production/annihilation, coulomb collisions, and prescriptions for particle heating/acceleration.

% The particle energy is described by the relativistic Lorentz factor $\gamma = \frac{E}{mc^2} = \sqrt{p^2+1}$,  with $p=\frac{\textit{P}}{mc}$ the specific momentum. The photon energies are described by their frequencies $\nu$. R is the typical length scale of the emission region and is one of the free parameter, in our study we consider a very small emission region with a radius R of 2 $r_G$. The time dependent kinetic differential equations are solved for the photons and the leptons. The code can also solve for the proton distribution. The code iterate the equations at each time step until the conservation laws (energy conservation for each radiative processes, particle number conservation) are satisfied to some specific precision.
% 
 \subsection{Radiative processes}
Synchrotron radiation is produced by charged particles spiraling around magnetic field lines. 
%It is one of the most important processes in astrophysics. Synchrotron emission and absorption both influence the involved particles by cooling and heating them respectively.
%The self-absorbed synchrotron radiation is characterized by the emissivity $j_s(p, \nu)$ (erg s$^{-1}$ Hz$^{-1}$) and absorption cross section $\sigma_s (p, \nu)$ (cm$^{2}$) and depends on the magnetic field, whose intensity is characterized by the magnetic compactness 
It depends on the magnetic field $B$ whose intensity is described by the magnetic compactness:
\begin{equation}
l_B=\frac{\sigma_T R}{m_e c^2}\frac{B^2}{8\pi}
\label{lb}
\end{equation}
%This magnetic field compactness is one of the free parameters of the model. The synchrotron expression for $j_s$ is taken from \citet{ghisellini98} and \citet{Katarzynski06b} for sub-relativistic particles, and from \citet{crusius86} and \citet{ghisellini88} in the relativistic regime. The expressions for $\sigma_s$ are computed analytically from $j_s$ (\citealt{leroux61, ghisellini88, ghisellini91}).
where $m_e$ is the electron mass, $c$ is the speed of light, $\sigma_T$ is the Thomson cross section, and $R$ is the size of the emission region. Synchrotron emission at frequency $\nu$ from a single electron with momentum $p$ is characterised by the emissivity $j_s(p, \nu)$ in erg s$^{-1}$ Hz$^{-1}$\citep{ghisellini88,ghisellini98, Katarzynski06b}. Synchrotron emission typically produces soft photons and cools the high energy emitting particles. The cooling time of relativistic particles emitting at frequency $\nu$ is:
\begin{equation}
t_{\rm synch}=1.29\times 10^{12} \times \nu ^{-1/2} \times B^{-3/2} \  (s)
\label{synch}
\end{equation}
For typical values of B, we have $$t = 1.3  \  (\nu/10^{18}\rm Hz)^{-1/2} (B/100 \rm G)^{-3/2} \ s $$ This corresponds to very short time scales, and only particles emitting at frequency lower than $10^{12}$ Hz cool on time scales comparable or longer than the duration of a typical flare (1000s). These particles are not observed to contribute much to the total emission. Moreover in our study we are not modeling the emission below $10^{12} Hz$ that is extended radio emission from the outflow.
Low energy particles can also absorb photons through the synchrotron process. Such absorption is described by the absorption cross section $\sigma_s(p,\nu)$ \citep{crusius86,ghisellini98}. The joint effect of high energy particle cooling and low energy particle heating tends to thermalize the particle distributions. It is called the {\it synchrotron boiler effect} \citep{ghisellini88}.

Photons of the emission region can also be scattered by Compton interactions. The scattering of isotropic photons of energy $h\nu _0$ by isotropic particles of energy $E_0=\gamma_0 m_e c^2$ is characterized by the resulting distribution of scattered photons $\sigma_c(p_0, \nu_0\rightarrow \nu)$. The {\sc belm} code uses the exact, Klein-Nishina cross section \citep{jones68,belmont09}. In the case of SgrA*, soft photons are up-scattered by high energy particles which brings them to high energy. This also cools the scattering particles. From equation (37) of \citet{piran04}, the typical inverse Compton cooling time is:
\begin{equation}
t_{\rm comp} =  3.1 \times 10^{10} \times \nu^{-1/4} \times B^{-7/4}  \ (s)
\label{compt}
\end{equation}
Again, this time scale is much shorter than the flare duration. 

%The differential Klein-Nishina cross section $\sigma_c(p_0, \nu_0 \rightarrow \nu )$ is used (\citealt{jones68, belmont09}). The photon frequency $\nu (p)$ is constrained by the energy conservation during one scattering event:
%\begin{equation}
%h\nu (p) - h\nu _0 + \gamma m_e c^2 - \gamma _0 m_e c^2 =0
%\label{compton}
%\end{equation}
%In the small angle scattering limit, when scattered photons and particles have energies similar to those incoming ($\Delta p(p_0, \nu_0 )/p_0 \ll 1$, and $\Delta \nu (p_0,\nu _0)/\nu_0 \ll 1$), a Fokker Plank approximation is used, otherwise the integral approach is used.
The effect of self-absorbed bremsstrahlung radiation is also computed. However for the inner most regions of the accretion flow, bremsstrahlung emission is a negligible component of the resulting spectra and it will not be discussed on this paper.

Photon-photon pair production and pair annihilation are also implemented in the code. However, we aim at modelling the emission from SgrA* only below 100 keV where these processes are negligible. They were disabled in order to reduce the computation time. 

Rather than computing the path of photons out of the emitting region with Monte Carlo simulations, photons produced in-situ are assumed to escape with a probability representative of the geometry. This probability depends on the photon energy.  For instance, high energy photons do not inverse Compton scatter and can escape freely when the optical depth is large, low energy photons can be scattered so much that they remain trapped in the system much longer. We use the escape rate from \citet{lightman87, coppi00} that reproduces well the results of Monte Carlo simulations in a spherical geometry. 
At low energy, synchrotron and bremsstrahlung can absorb photons before they escape. This modifies the escape probability in this energy range. We include the corresponding modifications to escape probability derived from \citet{sobolev57} \citep[see also][]{poutanen09}.

\subsection{Particle acceleration and heating}

The particle distribution depends on the above mentioned radiative processes and on several additional processes. 

Coulomb collisions tend to thermalize the particle distributions. The {\sc belm} code include Coulomb cross sections derived from \citet{nayakshin98}. However, the very low density inferred for SgrA* make this process very inefficient. In all the results shown in this paper, real Coulomb collisions are completely negligible. 

In order to account for the observed high energy radiation, particles needs to be heated/accelerated to high energy. Solving for the particle distribution thus requires to address also the physics of particle acceleration/heating.
Many processes have been proposed to account for high energy particles (viscosity, reconnection, shocks, first and second order Fermi processes,  etc.). However, the precise process at work in SgrA* is still unknown. Moreover the physics of these processes is not constrained well enough to have a precise modelling for their effect on the particle distribution. Only stochastic acceleration by MHD waves can be implemented easily in a Boltzmann equation for the particle distribution (see \citealt{liu06} for an application to SgrA*). However, if particle escape is slow, it forms a quasi-Maxwellian distribution and does not reproduce hard non-thermal distributions such as the one observed by NuSTAR. If particle escape is efficient, it can produce power-laws only if the accelerating rate has the same energy dependence as the radiative cooling, which is very unlikely \citep{katarzynski06}. 
%For difference dependencies, the produced distributions are complex and it could be addressed in a future work. 
Therefore we use very general, ad-hoc prescriptions, inspired from what is done for the corona of accreting black holes (such as {\sc Eqpair}, \citealt{coppi00} ; or {\sc belm}). 
%I.e. we mimic thermal processes (such as viscosity, or collisions with hot protons) using virtual collisions with virtual protons of high energy, and we mimic non-thermal processes by taking particles from all energies and re-injecting them as a power-law distribution. This heating and acceleration compete with all radiative processes to produce complex distributions of particles. 
%Particle acceleration can be done in different ways: instantaneous injection of energetic particles, power provided to the electron distribution by some unspecified process, Coulomb-like heating (heating by thermal protons but with enhanced efficiency), second order Fermi acceleration. In our study we will explore the two first possibilities, equivalent to assuming that no shocks are present. 
 We use two different channels to provide energy to the particles.
\begin{itemize}
\item We mimic thermal processes by computing Coulomb collisions with a virtual population of hot protons (with temperature $k_BT_p = 40 $ MeV). Real collisions are very inefficient and do no provide significant heating, whatever the proton temperature. Rather, this prescription aims at reproducing the effect of anomalous processes (such as viscosity) on the lepton distribution. Therefore, the heating efficiency is renormalised by an arbitrary constant so that to inject power $L_{\rm th}$ (erg s$^{-1}$) into the emitting region. In the following, this free parameter will be described by the compactness parameter $l_{\rm th} = \sigma_TL_{\rm th}/(Rm_ec^3)$. Such prescription not only heats the global distribution of particles. It also thermalises it. As the efficiency of the virtual collisions is enhanced, the efficiency of the  associated thermalisation is also enhanced to an anomalous level. Anomalous heating is a common feature of accreting systems, so such heating is not surprising even though the origin is debatable.

\item We model non-thermal processes by constantly injecting particles with a power-law distribution $N(\gamma)\propto \gamma^{-s}$. This distribution is characterised by 4 parameters: the slope $s$, the minimal and maximal energies $\gamma_{\rm min}$ and $\gamma_{\rm max}$ respectively, and the normalisation. As we want this process to keep the number of particles constant, the re-injected particles are taken from the lepton population itself, with a uniform probability, independent of their energy. In the following, the minimal energy of the power-law will be set to $\gamma_{\rm min} = 50$ , so that particles are accelerated from the bulk of the distribution. Indeed, the thermal peak of the spectrum (around $10^{12}$ $Hz$) imply an electron temperature around $10^{11}$ $K$. And the maximal energy of accelerated particles will be set to $\gamma_{\rm max}= 4.6 \times 10^5$, large enough to reproduce the NuSTAR data. The high energy cutoff has not been confirmed by NuSTAR observations (Barri\`ere et al., submitted), and the possible physical processes responsible for the non-thermal component (turbulent acceleration, reconnection, weak shocks) can accelerate electrons to very high energies. The normalisation is computed so that the non-thermal  process injects into the region a power $L_{\rm nth}$, described by the free parameter $l_{\rm nth} =\sigma_TL_{\rm nth}/(Rm_ec^3)$. The slope is also a free parameter of the model. 
\end{itemize}
Such prescriptions compete with all other processes to produce complex distributions of particles.

\subsection{Modelling the particle dynamics}

%The code allows the injection of particles. This injected population can come for instance from the surrounding accretion flow. Any distribution $\dot{N}^{inj} _e$ can be injected at each time step (thermal, Gaussian, power-law, mono-energetic, etc.). The injection of particle is controlled by the particle injection compactness, which is another free parameter:
%\begin{equation}
%l_{inj}  = \frac{\dot{E}_{inj}}{m_e c^3 R/\sigma _T}=\frac{R^2 \sigma _T}{c}\frac{4\pi}{3}\int \gamma \dot{N}^{inj}_e dp
%\label{ParticleInjCompactness}
%\end{equation}
%similarly, the code allows photon injection controlled by the parameter:
%\begin{equation}
%l_{inj,\nu}= \frac{L_{inj}}{m_e c^3 R/\sigma _T}=\frac{R^2 \sigma _T}{c}\frac{4\pi}{3}\int \frac{h\nu}{m_ec^2} \dot{N}^{inj}_\nu d\nu
%\label{PhotonInjCompactness}
%\end{equation}

Although observational evidence clearly indicates that the sub-mm to X-ray emission originates from a very small region close to the black hole, the dynamics of the particles is very uncertain. Free, relativistic particles can travel though the emitting region in a light crossing time:
\begin{equation}
t = R/c = 43 \ s
\end{equation}
which is much shorter than the flare duration. 

However, the medium is magnetised so that particles are not free to move on straight trajectories. Instead, they are bound to the magnetic field lines. For magnetic intensity of 1-100 G, even X-ray emitting particles have gyro-radii orders of magnitude smaller than the emitting region. Therefore, if the medium is turbulent and the magnetic field tangled, even the highest energy particles can be considered as trapped in the main flow.

The detailed structure of the accretion flow and in particular the accretion velocity are not known. Therefore we investigate two extreme scenarios. 

\subsubsection{The closed system approximation}

On the one hand we consider that the accretion velocity is very small. In that case, particles remain a very long period of time in the emitting region. Radiative and acceleration processes set steady distributions of particles and spectra before particles escape from the system. 

This model is characterised by 5 free parameters:  the lepton density $n_e$ related to the Thomson optical depth $\tau$ by $\tau = n_e \sigma_T R$, the magnetic field compactness $l_B$ defined by Eq. \ref{lb}, the power of the thermal heating and non-thermal acceleration characterized by the compactness parameters $l_{\rm th}$ and $l_{\rm nth} $ respectively, and the slope of the non-thermal heating process $s$.
In this model without particle escape, the particle distribution results from the balance between thermal heating, non-thermal acceleration, and radiative cooling.
At high energy, thermal heating is inefficient, and the particle distribution results from the balance between radiative cooling and non-thermal acceleration. In good approximation, Compton and synchrotron processes have the simple cooling laws shown in Eq. \ref{compt} and \ref{synch}. As acceleration tends to produce an electron power-law distribution of index $s$, the steady distribution is also a power-law with index $s'= s +1$. When synchrotron radiation is the dominant process, this produces a synchrotron spectrum of spectral index $\alpha = s' /2$. At lower energy, the physics and the shape of the particle distribution are more complex.

We solve this model numerically for different parameter sets presented in Figure \ref{Quiescent00}, \ref{Flare00c}, and \ref{Flare00s}.

%The code allows for the escape of particles with escape probability proportional to their velocity $P^{esc}_e = \frac{\beta c}{R}$. Other escape laws can also be defined.\\
%High energy photons do not inverse Compton scatter and can escape freely when the optical depth is large, low energy photons can be scattered so much that they are trapped in the system. We use the escape rate $r^{esc}_\nu = P^{esc}_\nu\frac{c}{R}$ from \citet{lightman87} that reproduces well the results of Monte Carlo simulations in a spherical geometry.

\subsubsection{The open configuration}

On the other hand, we also consider the extreme case where matter flows in and out of the emitting region with an accretion velocity approaching the speed of light. In that case, particles escape the emitting region on time scales comparable to the light crossing time, i.e. comparable also to the radiative times scales. Escape can therefore compete efficiently with radiation and acceleration processes. This model will be referred to as the {\it open configuration}. The distribution of the matter entering the emitting region needs to be given $\dot{N}_{\rm inj}(\gamma)$. We assume that non-thermal acceleration occurs only in the emitting region, so that particle entering this region have a thermal distribution described only by 2 parameters: its temperature $\theta_{\rm inj} = k_BT_{\rm inj}/(m_ec^2)$ and its normalisation. The former is set to $\theta_{\rm inj}= 13$ in all models. The latter is described by the injection compactness:
\begin{equation}
l_{\rm inj}  =\frac{4\pi}{3} \frac{R^2 \sigma _T}{c}\int \gamma \dot{N}_{\rm inj} d\gamma \approx 3 \theta_{\rm inj} \frac{R^2 \sigma _T}{c} \dot{n}_{\rm inj}
\label{ParticleInjCompactness}
\end{equation}
where $\dot{n}_{\rm inj}$ is the total number of particles injected into the emitted region per unit time, and the last equality holds for thermal distributions with relativistic temperature ($\theta_{\rm inj} >> 1$). Once in the emitting region, particles are assumed to escape on a typical time scale $t_{\rm esc}=R/c$, which corresponds to an escape  probability $p_{esc} = R/(c \ t_{\rm esc}) = 1$. 
%Other escape laws can also be implemented such as an escape probability proportional to the particle velocity ($p_{esc} =\beta R/c$ for instance), but a constant escape probability is more consistent in our case as we consider an open region. 

This model is described by 4 free parameters: the magnetic field compactness $l_B$, the non thermal compactness $l_{\rm nth}$ and the slope of the power-law $s$, and the injection compactness $l_{\rm inj}$. 
%and the temperature $\theta_{\rm inj}$ of the incoming distribution.  
The particle density is no longer a free parameter and results from the balance between injection and escape. When injected particles have a relativistic temperature, the steady-state optical depth is:
\begin{equation}
\tau_{T}=\frac{1}{4\pi}\frac{l_{\rm inj}}{\theta_{\rm inj}}
\label{Tau}
\end{equation}

In that model, the steady particle distribution results from the balance between injection and non-thermal acceleration versus both escape and radiative cooling. In this configuration, the shape of the steady state distribution is more complex than in the closed system.  At high energy, particles cool before they escape. As in the closed system, the leptons form a power-law distribution of index $s' = s +1 $, and emit a power-law synchrotron spectrum of spectral index $\alpha = s' /2$. At low energy, particles escape before they cool and the steady state electron distribution is a power-law of slope $s'=s$. This produces a synchrotron spectrum of spectral index $\alpha = s/2$. The particle distribution and photon spectrum thus exhibit a break, whose energy depends on the relative efficiency the cooling and escape. As far as synchrotron radiation is the dominant cooling process, the break in the photon spectrum is:
  \begin{multline}
\nu _{break}=2.97 \times 10^{14} \left( \dfrac{s-1}{3.6 -1}\right)^{-2}\left(\dfrac{T_{esc}}{R/c}\right)^{-2}\\ \left(\dfrac{R}{3 \times 10^{12}}\right)^{-2}
\left(\dfrac{B}{50G}\right)^{-3}  \ \rm Hz
\label{CoolingBreak}
\end{multline}
Such a model was for instance proposed by \citet{doddseden10} to explain the flaring X-ray emission without violating the NIR upper-limits. They used a broken power-law distribution. However, depending on the parameters, processes other than synchrotron emission can contribute to the physics. Also, a self consistent cooling break is not sharp and extend over a significant frequency range. Here we extend their conclusion by solving self-consistently for the particle distribution. 

%\subsection{Time scale and modelling of the system}
%We model the emitting region in the single-zone approximation, which corresponds to the physics of a homogeneous and isotropic sphere consisting of mater and photons, representative of a small region of the plasma surrounding Sgr A*. We set the radius of the sphere to be R = 2 $\rm r_G$ in all our models based on the size derived from the flare time scale variability.  A description of the characteristics of our different models are summarized in table 1. 
\begin{table*}
 \centering
 \begin{minipage}{170mm}
  \caption{\textit{Description of simulations}}\footnotetext[0]{\textit{The first column gives the Figure's number. The second  one gives the state of the spectrum we are trying to reproduce (in quiescence or during a flare) and the configuration of the model (closed or open region). $B$ is the magnetic field which is a free input parameter. $n_e$ is the density which is also an input parameter in the closed configuration cases (Figure \ref{Quiescent00}, \ref{Flare00c}, and \ref{Flare00s}) but an output of the simulations in the other cases. $l_{\rm th}$ is the prescription for thermal heating, and $l_{\rm nth}$ is the prescription for non-thermal heating/acceleration . $l_{\rm inj}$  is the injection of particles for the open configuration cases. $s$ is the spectral index used in the non-thermal prescription for acceleration. $P_{esc}$ gives the probability for particles to escape from the system. $\epsilon_k/\epsilon_b$ is the ratio of kinetic energy over magnetic energy ($8\pi \int mc^2(\gamma - 1)N_{\gamma}d_{\gamma} / B^2$) and is computed after the different models have run. $\rm T_e$ is the temperature of the thermal part of the spectrum. Because our lepton distributions are calculated self-consistently we never have a perfect thermal distribution, but this temperature corresponds to the closest match between the real distribution and the pure Maxwellian (plotted in dotted lines in all our distributions as a comparison), this is of course an output of our models as well. Finally, the last column gives the total luminosity of the spectrum. In the closed configuration case, this luminosity is the direct result of the thermal plus non-thermal compactness parameters, while in the open configuration it results from the balance between injection, escape, and cooling. We remind that in all our models we have  $\rm R=2 r_G$, $\gamma_{min} = 50$, $\gamma_{max}=4.6\times 10^5$, and $\theta_{inj}=13$.} }
  \begin{tabular}{@{}lcccccccccccc@{}}
  \hline
Model  & spectrum & configuration & B  &  $n_e$  &  $l_{\rm th}$  &  $l_{\rm nth}$ &  $l_{\rm inj}$  &  $s$ & $p_{esc}$ & $\epsilon _k/\epsilon _b$ & $\rm T_e$&$L_{bol}$  \\ 
  &  &  &\tiny{Gauss}  &  \tiny{part/$cm^3$}  & \tiny{$\times 10^{-5}$}   &  \tiny{$\times 10^{-5}$} &   \tiny{$\times 10^{-4}$} &   &  &  & \tiny{$\times 10^{10}$ K} & \tiny{$\rm \times 10^{36} erg \ s^{-1}$} \\ 
\hline
  Figure \ref{Quiescent00} & \tiny{Quiescent} & \tiny{Closed} & 154.3 & $4.6\times 10^6$ & 2 & 1 & 0 & 3.60 & 0 & 0.15 & 11.4 & $1.4$\\
  Figure \ref{Flare00c} & \tiny{Flare} &  \tiny{Closed} & 48.8 & $4.6\times 10^6$ & 7& 3 & 0 & 2.40 & 0 & 6.32 & 35.0 &$4.8$\\
 Figure  \ref{Flare00s} & \tiny{Flare} & \tiny{Closed} & 48.8 & $4.6\times 10^6$ & 5 & 4 & 0 &  2.20 & 0 & 5.79 & 33.4 &$4.3$\\
  Figure \ref{Quiescent33} & \tiny{Quiescent} & \tiny{Open} & 175.3 & $3.3\times 10^6$ & 0 & 1 & 4.64 &  3.60 & 1 & 0.08 &  7.84 &$1.0$\\
  Figure \ref{Flare33s_J21} & \tiny{Flare} & \tiny{Open} & 175.3 & $3.3\times 10^6$ & 0 & 9.8 & 4.64 & 2.28 & 1 & 0.09 & 7.90 &$3.8$\\
  Figure \ref{Flare33s_O17} & \tiny{Flare} & \tiny{Open} & 175.3 & $3.3\times 10^6$ & 0 & 11.4 & 4.64 &  2.13 & 1 & 0.09 &  7.87 & $4.8$\\
  Figure \ref{Flare33c} & \tiny{Flare} & \tiny{Open} & 34.5 & $1.4\times 10^8$ & 0 & 100 & 200 &  2.60 & 1 & 97.16 &  7.79 &$5.8$\\
  Figure \ref{Quiescent33_x3} & \tiny{Quiescent} & \tiny{Open} & 175.3 & $9.9\times 10^6$ & 0 & 1 & 14 & 3.60 & 1 & 0.24 &  7.86 & $3.8$ \\
\hline
\end{tabular}
\end{minipage}
\end{table*}

\section{Sgr A* resulting spectra}

The flare duration varies but the typical time is about 3000 seconds. The radiative time scales and the thermalisation time are much smaller than the flare duration. The particle and photon distributions are in quasi steady state at each moment of the flare, and the flare evolution is directly governed by the evolution of model parameters, here namely the acceleration processes described here by the parameters $\rm l_{nth}$ and $\rm l_{th}$.   Therefore we will mostly present and discuss the results from steady state simulations. I.e. we aim at reproducing separately the quiescent and flaring spectra by changing the value of only few parameters. Doing so, we can say that the flaring state is just a transition governed by the increase or decrease of few physical parameters. What is driving these changes is subject to interpretation, but knowing what needs to be modified should give us some good first insight into the physical processes at work.

Table 1 summarises the characteristics of our different models. 

\subsection{Data}

Sgr A*'s SED is made of different observations that are variable and in most cases have not been observed simultaneously in different wavelengths. We have chosen a set of data representative of the overall variation of the spectrum. In all figures, the black radio points are from \citet{falcke98} and \citet{zhao03}, the red radio points are recent ALMA observations from Brinkerink, Falcke et al. submitted). The black far IR upper limits are from \citet{serabyn97} and \citet{hornstein02}, the green MIR data are from \citet{schodel11}, the pink NIR  lower point (in the quiescent spectra) and the cyan NIR upper point (in the flare spectra) are from \citet{ghez04, genzel03}, and \cite{doddseden11}. The green ``bowtie'' is from \citet{bremer11} and is one of the few slopes that has been observed so far in the IR. Several flare observations seem to be consistent with this value of the NIR spectral index around $-0.6 \pm 0.2$ \citep{ghez05,gillessen06,hornstein07,bremer11}. 

The black ``bowtie'' in the quiescent X-ray is from \cite{baganoff01, baganoff03} and is an upper limit for the central emission because it is contaminated by thermal bremsstrahlung from the outer accreting matter.  The orange ``bowtie'' is a Chandra flare from \citet{nowak12}. Finally the blue data points (dark and light blue) are two flares observed with NuSTAR on July 21st and October 17th 2012 respectively \citep{barriere14}. Even-though the X-ray flares may seem different in shape and slope, they are both acceptably fit by an absorbed power-law, and their photon indices are not significantly different ($2.23^{+0.24}_{-0.22}$ and $2.04^{+0.22}_{-0.20}$ for the July 21st and October 17th flares, respectively). Barri\`ere et al. 2014, investigated the presence of a cutoff in the October 17th flare, but found that it is not required by the data. The other wiggles in this spectrum (one may see a "V" shape in the low energy part of the spectrum) are not significant either. One need to keep in mind that the error bars show the 1-sigma confidence range, which means that an acceptable fit does not need to go through them all but 3 out of 9. 

Three X-ray flares are shown on the first flaring spectra (Figure \ref{Flare00c} and \ref{Flare00s}), but then we consider only the modelling of one of the NuSTAR flares that extend to higher energies. For a better reproduction of the spectrum, we need to move to the ``open configuration'' where we present some possible spectra for the July flare or the October flare. 

%Three X-ray flares are shown on the first flaring spectrum (Figure \ref{Flare00c}), but then we consider only the modelling of one of them, namely the dark blue NuSTAR data set, which is a flare observed in July 2012, as we also model the light curve for this specific flare event.

\subsection{Sgr A* spectra from a closed region}

\begin{figure}
 \includegraphics[scale=0.35]{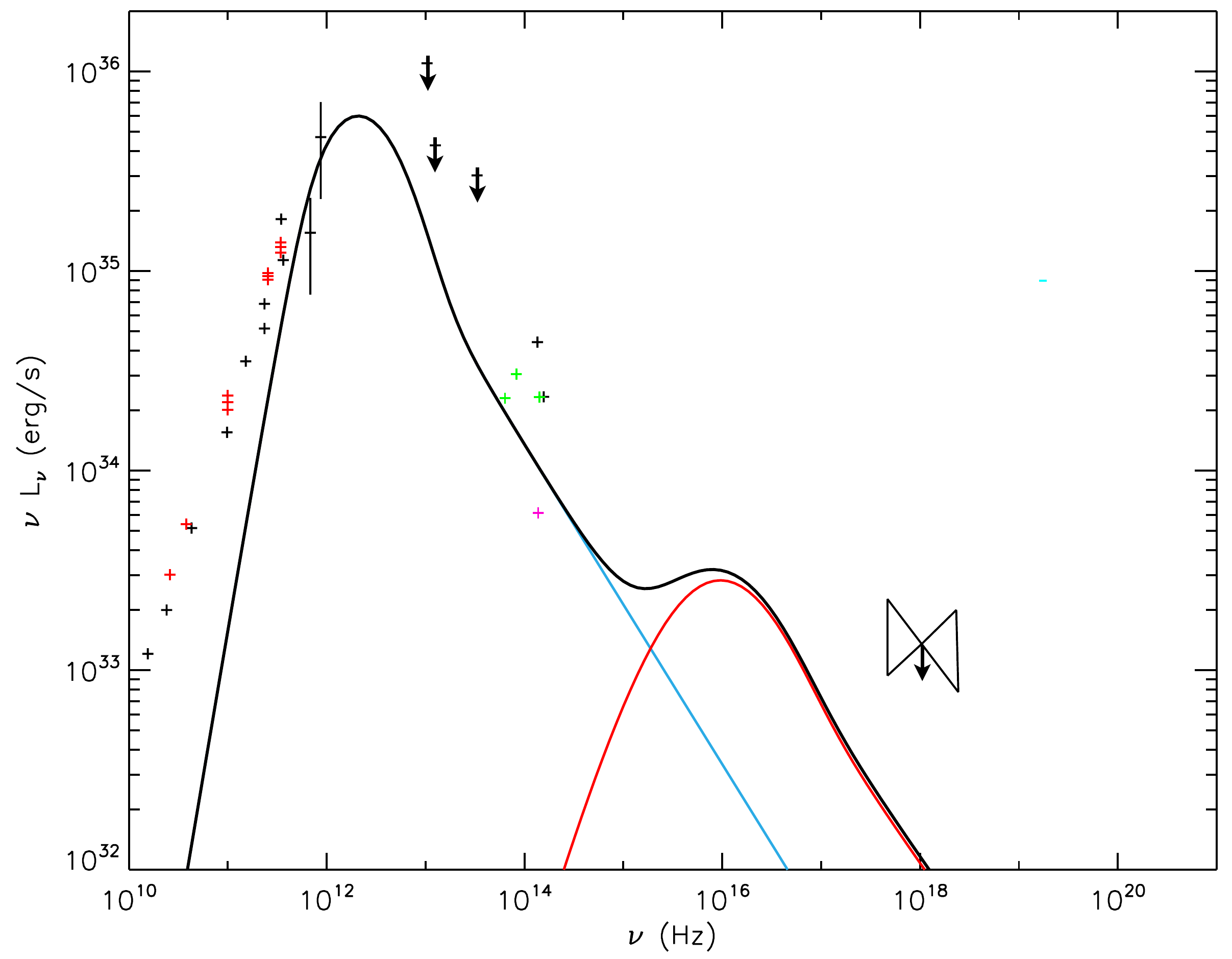}
 \includegraphics[scale=0.35]{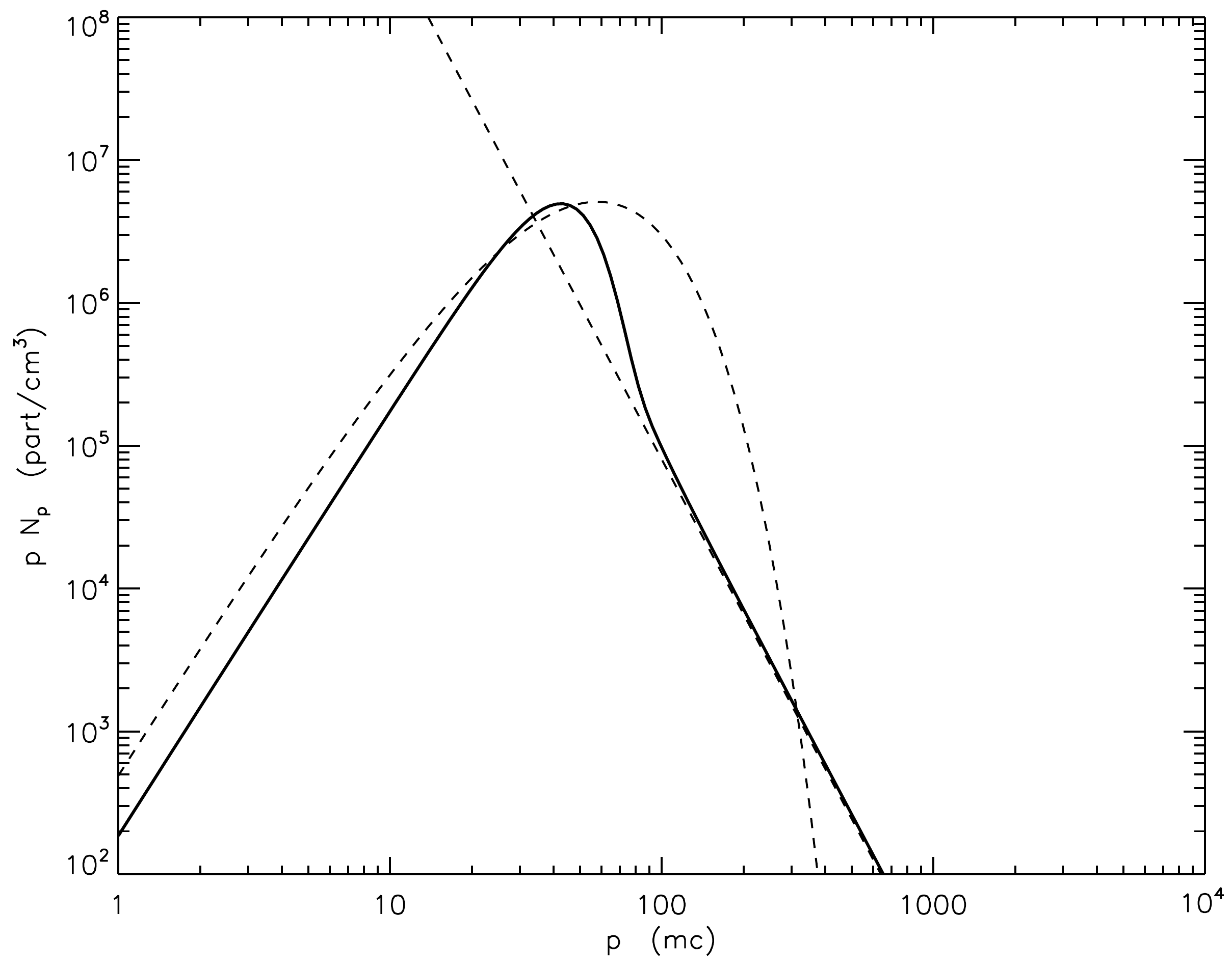}
 \caption{\textit{Quiescent spectrum from Sgr A* (top panel) and the associated lepton distribution (bottom panel) in a closed system configuration. On the spectrum, the black data points are taken from \citet{yuan03} with the X-ray ``bowtie'' corresponding to an upper limit for the quiescent state of Sgr A*. The data points on the spectrum are described in details in section 3.1. The blue curve component of the spectrum corresponds to the synchrotron process, while the red one corresponds to the Compton process. The Bremsstrahlung contribution is not visible in the scale of this plot. On the electron distribution, the solid line is the shape of the calculated distribution from which the spectrum comes from, while the dotted lines indicate a pure Maxwellian plus power-law components for comparison.}} 
\label{Quiescent00}
\end{figure}

\begin{figure}
 \includegraphics[scale=0.35]{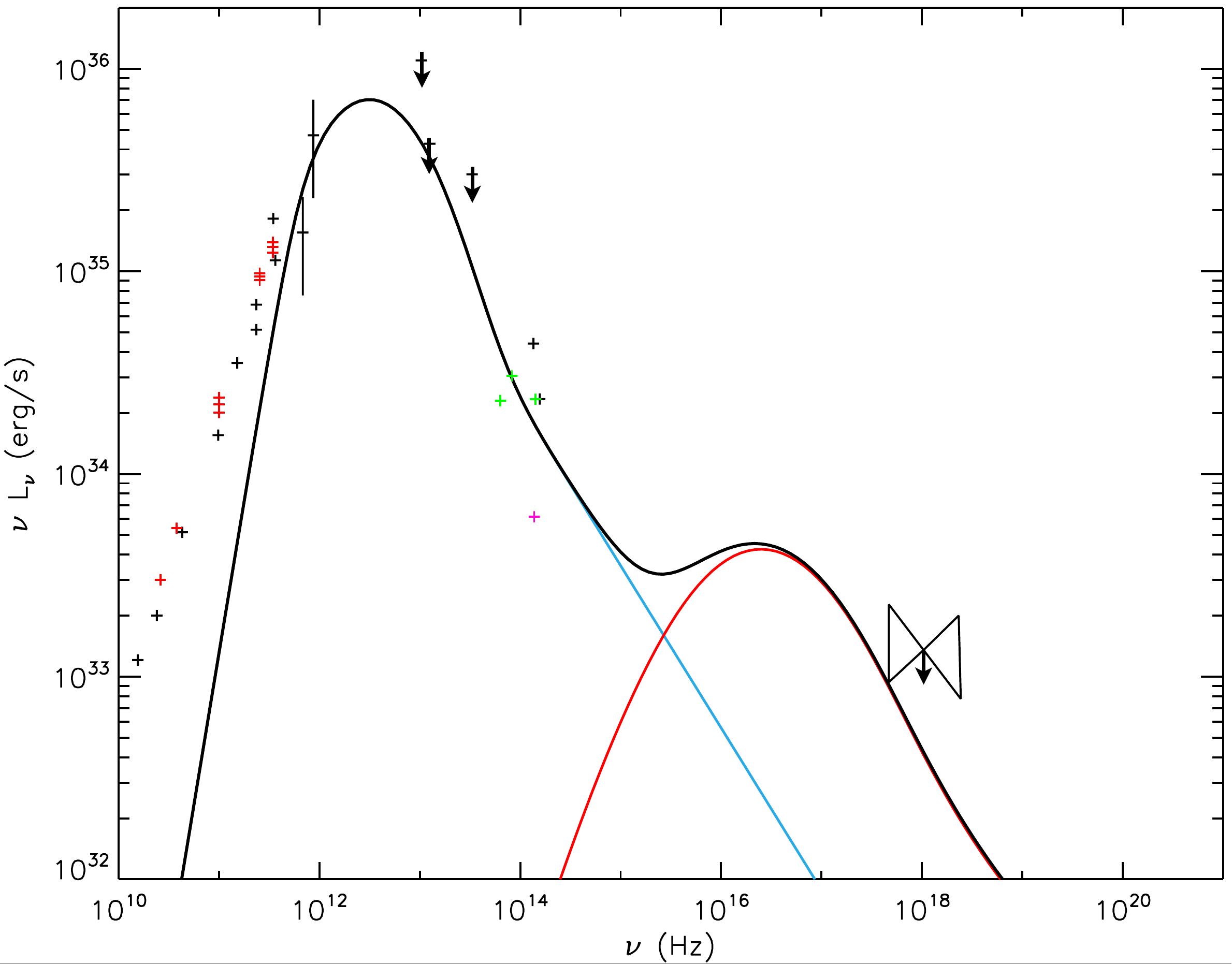}
 \caption{\textit{Quiescent spectrum from Sgr A* resulting from the``standard'' distribution consisting of a simple Maxwellian plus power-law (dotted line in the bottom panel of Figure \ref{Quiescent00}). }} 
\label{Fix}
\end{figure} 

Figure \ref{Quiescent00} shows a spectrum for the quiescent state of Sgr A*, together with the emitting steady state lepton distribution.  For this first spectrum, we consider a density of $4.6\times 10^{6}$ particles per cubic centimetre (which corresponds to $\tau=4 \times 10^{-6}$) and we keep this density to study the closed region, i.e. we consider that the number of particles is kept constant in the quiescent and flaring states. In quiescence, the magnetic field is 154.3 Gauss, the plasma is magnetically dominated with $\epsilon _k/\epsilon _b=0.15$.  The thermal heating is twice the non-thermal one and corresponds to $9.6\times 10^{35}$ erg $\rm s^{-1}$ and $4.8\times 10^{35}$ erg $\rm s^{-1}$ respectively, so that the total emission reaches $1.4 \times 10^{36}$ erg $\rm s^{-1}$ in quiescence.  We can see in the resulting spectrum in Figure \ref{Quiescent00} that the non-thermal component is not dominant, and this is not only due to the low value of $\rm l_{nth}$ but also because of the steep injected slope $s=3.6$. Nevertheless, this non-thermal component is important in order to reproduce the lower NIR data point. The thermal part contributes mainly to the sub-millimetre bump. In this case, we can notice how the thermal part of the lepton distribution differs from the standard Maxwellian shape on the bottom panel of Figure \ref{Quiescent00}. Indeed, for particle energies around $ p=10^2 mc$, the difference between the calculated distribution and the standard shape in dotted line, can reach almost two orders of magnitude. The steady state particle distribution is sharper than a pure Maxwellian. Above $\gamma = 100$, synchrotron cooling overcomes the anomalous thermalisation, and the distribution cuts more sharply than a pure thermal one. This also produces a sharper sub-mm bump as is illustrated in the resulting spectrum (top panel of Figure \ref{Quiescent00}). On Figure \ref{Fix} we plotted the spectral shape resulting from the dotted line of Figure \ref{Quiescent00}. We can see that the quiescent spectrum would be much wider, reaching the far-IR upper limits. The novelty of our work is illustrated by the difference between Figure \ref{Quiescent00} and \ref{Fix}, that results from the careful and detailed calculation of the lepton distribution.

\begin{figure}
 \includegraphics[scale=0.35]{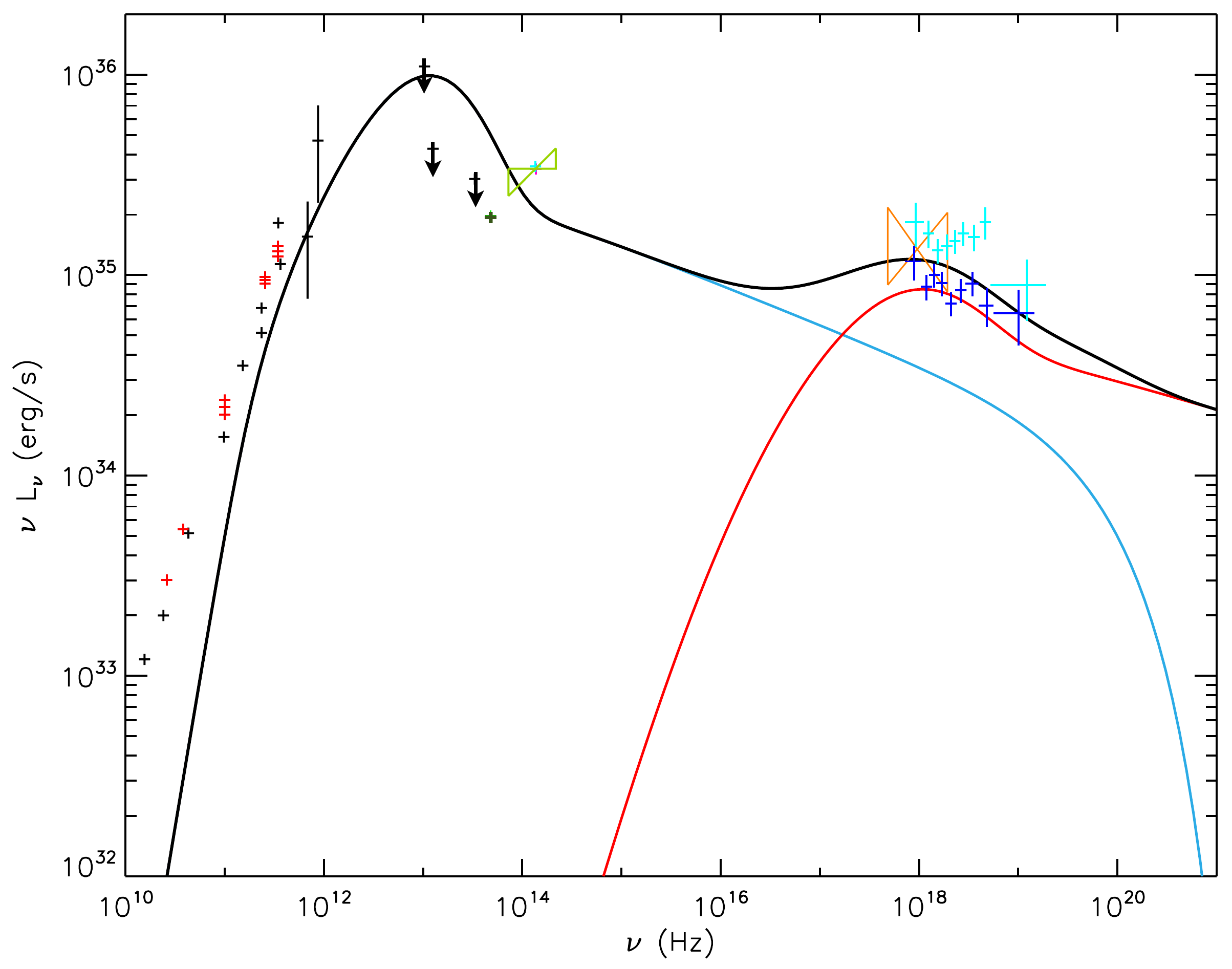}
 \includegraphics[scale=0.35]{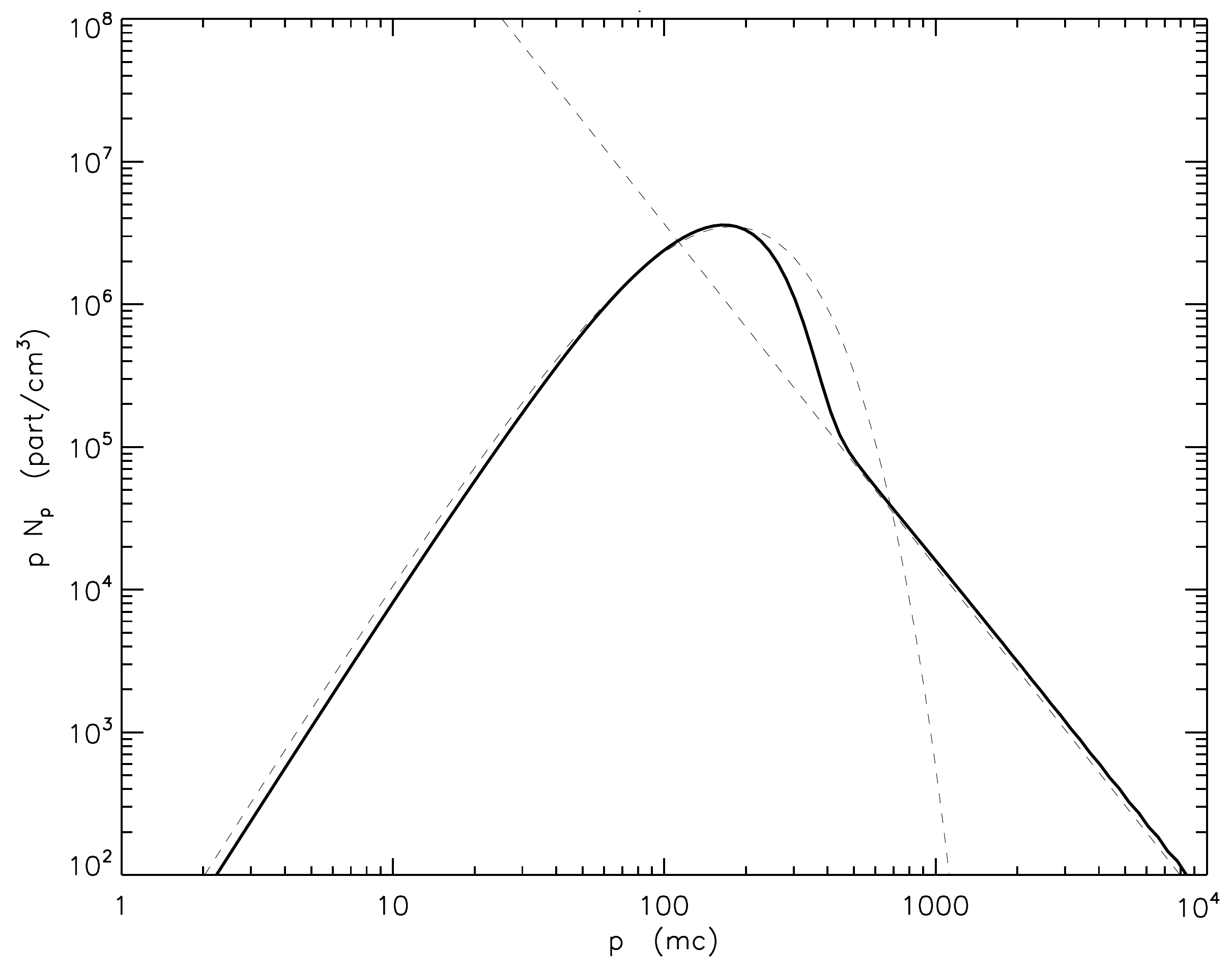}
 \caption{\textit{Flare spectrum from Sgr A* (top panel) and the associated lepton distribution (bottom panel) in a closed system configuration. The data points on the spectrum are described in the beginning of section 3. The blue curve corresponds to the synchrotron process, while the red corresponds to the Compton processes. The Bremsstrahlung contribution is too small to be visible on this scale. 
 On the lepton distribution, the full line corresponds to the actual calculated distribution, while the dotted line is a standard Maxwellian plus power law as a comparison. In all our models (except for Figure \ref{Fix}), the spectra result from the calculated particle distribution (full line), while the theoretical fixed distribution (dotted line) is just shown as an illustration.}} 
\label{Flare00c}
\end{figure}

Starting from similar conditions as in the quiescent state of Figure \ref{Quiescent00}, Figure \ref{Flare00c} shows a spectrum for the flaring state of Sgr A*, together with the lepton distribution. The spectrum is dominated by synchrotron self-Compton emission (red line), even-though the non-thermal synchrotron (blue line) has a non negligible contribution to the total spectrum. The emitting region is the same as in the quiescent state with the same density of particles. However the magnetic field has dropped from 154.3 to 48.8 Gauss, the plasma being now kinetically dominated with $\epsilon _k/\epsilon _b=6.3$. This dramatic change could be interpreted as being due to magnetic reconnection, a physical process that could be at  the origin of the flaring event. When the magnetic field is rearranged, some magnetic energy is converted to kinetic energy, thermal energy, and particle acceleration. In this way we have a decrease of the magnetic field strength and an increase of the two parameters $\rm l_{th}$ and $\rm l_{nth}$ representing the thermal and non-thermal acceleration respectively. In this case, to have the Compton dominated spectrum in the flaring state (Figure \ref{Flare00c}), the thermal heating increases from $9.6\times 10^{35}$ erg $\rm s^{-1}$ to $3.4\times 10^{36}$ erg $\rm s^{-1}$, and the non-thermal one from $4.8\times 10^{35}$ erg $\rm s^{-1}$ to $1.4\times 10^{36}$ erg $\rm s^{-1}$. The non-thermal component has also a much flatter distribution in the flaring state, meaning that the high energies are more populated, while $s$ is steeper during quiescence. With this model, during the flare, the total luminosity reaches $4.8 \times 10^{36}$ erg $\rm s^{-1}$.

\begin{figure}
 \includegraphics[scale=0.35]{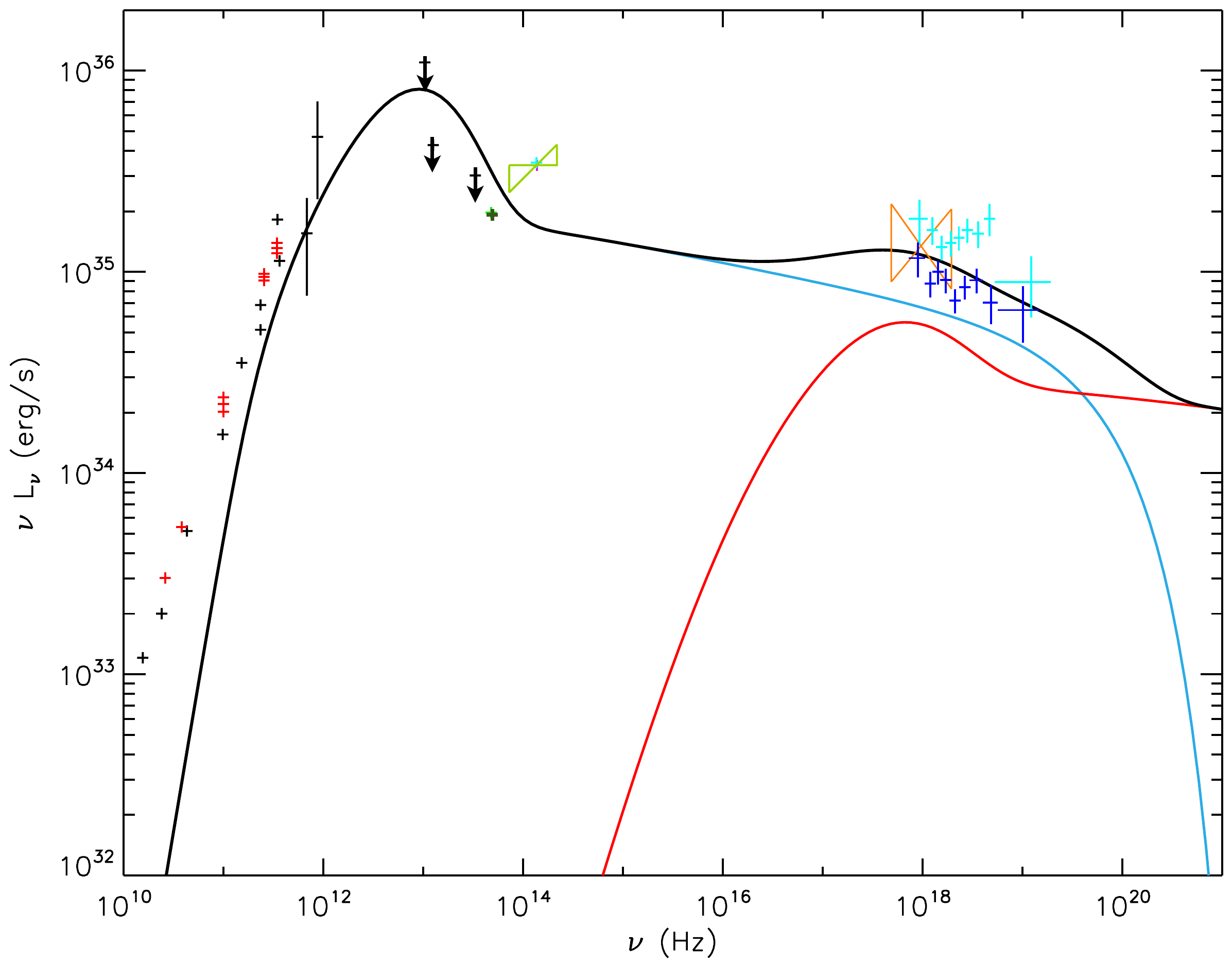}
 \includegraphics[scale=0.35]{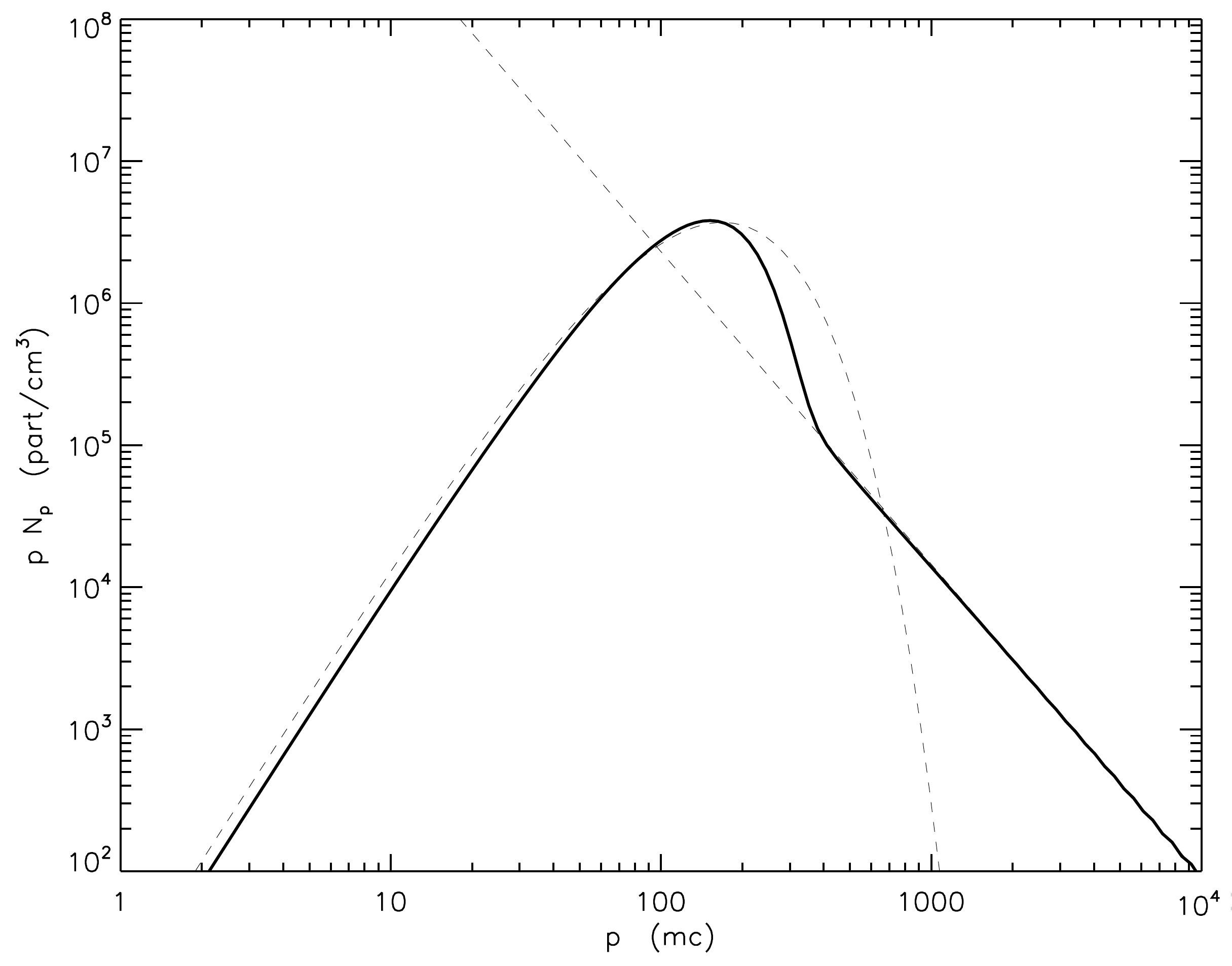}
 \caption{\textit{Flare spectrum from Sgr A* (top panel) and the associated lepton distribution (bottom panel) in a closed system configuration.  The data points on the spectrum are described in the beginning of section 3. The blue curve corresponds to the synchrotron processes, while the red corresponds to the Compton processes. It is the same as Figure \ref{Flare00c}, but for the case of a dominant synchrotron component with respect to the Compton one. On the electron distribution, the full line is the shape of the actual distribution from which the spectrum comes from, while the dotted lines are pure Maxwellian plus power-law components as a comparison.}} 
\label{Flare00s}
\end{figure}

Figure \ref{Flare00s} shows another potential flare model, together with its lepton distribution. This spectrum is similar to Figure \ref{Flare00c} except that the non-thermal synchrotron component is more important than the Compton one. The radius of the emitting region is the same as well as the density of particles. The magnetic field magnitude is also 48.8 Gauss. The difference comes from the balance between thermal versus non-thermal heating. For this non-thermal synchrotron dominated spectrum, the non-thermal contribution $l_{nth}$ is more important than previously with a value of $1.9\times 10^{36}$ erg $\rm s^{-1}$ and a slope of 2.2. The total luminosity is similar to the previous case with L=$4.3 \times 10^{36}$ erg $\rm s^{-1}$. 

%\subsection{Discussion}

The quiescent state is very well reproduced by this closed region configuration model, the sub-millimetre bump is clearly fitted by synchrotron emission which extends to the lower part of the variable MIR and NIR data, and we are not violating the X-ray limit represented by the black ``bowtie''. According to the new results from \citet{wang13}  and \citet{neilsen13} saying that the inner region is dominated by non-thermal emission from combined weak flares that can contributes to 10$\%$ of the quiescent X-ray, we could be even too low in the X-ray luminosity (lower than 10$\%$ of the observed flux). The flaring spectra  are somewhat more marginal because the sub-millimetre contribution is too high compared to two black data points that are upper limits and end up below the spectra. On the other hand, the sub-mm part of the spectrum is also variable on the order of 20$\%$ and considering that we don't have perfect simultaneous data, it still provide a close enough interpretation, meaning that we are still within 20$\%$ of the actual data point values. In the MIR, the flaring spectra shown in Figures \ref{Flare00c} and \ref{Flare00s} are in the right range of luminosity, and we can also reproduce the X-ray flare fluxes. The NuSTAR (blue) and Chandra (orange) flare slopes are respected, while the trend of one of the MIR flare (green ``bowtie'') is not well reproduced at all. We have to keep in mind that our data are not simultaneous and slopes in the MIR have been observed only few times, but still in this case our slope seems to be in contradiction with this observation.

Comparing models with observations, the  ``$\alpha$'' prescription from \citet{shakura73} is still used to parametrize turbulence and quantify the angular momentum loss mechanism, and the process whereby gravitational binding energy is converted into radiation.  The best physical interpretation of this $\alpha$ parameter is given by the mechanism of magneto-rotational instability (MRI). For instance, \citet{hawley95} have performed three-dimensional magneto-hydrodynamic numerical simulations of an accretion disk to study the nonlinear development of the MRI, they obtained that the time average of $\alpha$ is 0.6 for the vertical field runs. In the study of advection-dominated accretion and black hole event horizon, \citet{narayan08} argued that for an advection-dominated accretion flow,  the theoretically expected value of $\alpha$ is 0.1-0.3. We have the estimated numbers for $\alpha$ for the different cases studied here. 
 Assuming  Keplerian assumptions, which is obviously a very simplistic approximation close to the black hole but allows us to check roughly that the orders of magnitude are not inconsistent with the first order $\alpha$ approximation, the viscous heating $Q$ is related to the ``$\alpha$'' parameter by:
 \begin{equation}
 Q=\frac{3}{2}\alpha P \left(\frac{GM}{R^3}\right)^{1/2}
 \end{equation}
where P is the pressure. Using the dimensionless constants from our model, the viscous parameter is derived from the following formula:
\begin{equation}
\alpha =\frac{1}{2}\frac{l_{th} \ r^{1/2}}{\tau \ \Theta _e}
\end{equation}
with r=2$r_G$, $\tau = n_e \sigma _T R$, $\Theta _e=kT_e/m_ec^2$. We find that the viscosity parameters $\alpha$ resulting from models \ref{Quiescent00}, \ref{Flare00c}, and \ref{Flare00s} are 0.18, 0.18 and 0.16 respectively. These values are in good agreement with the theoretical predictions described above which illustrates that an anomalous thermal heating is common on the context of accretion disks.

\subsection{Plasma with escape and thermal injection}

\begin{figure}
 \includegraphics[scale=0.35]{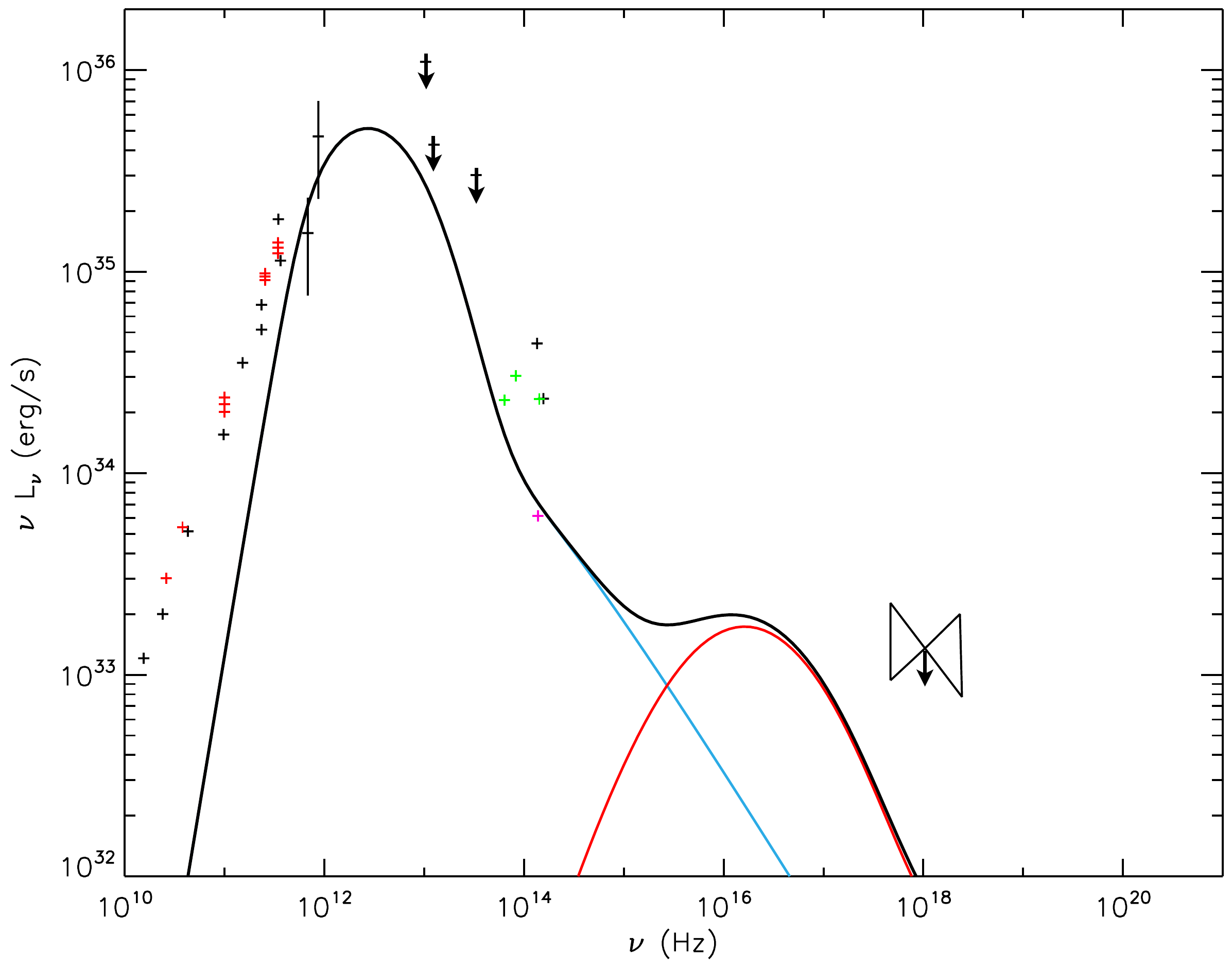}
 \includegraphics[scale=0.35]{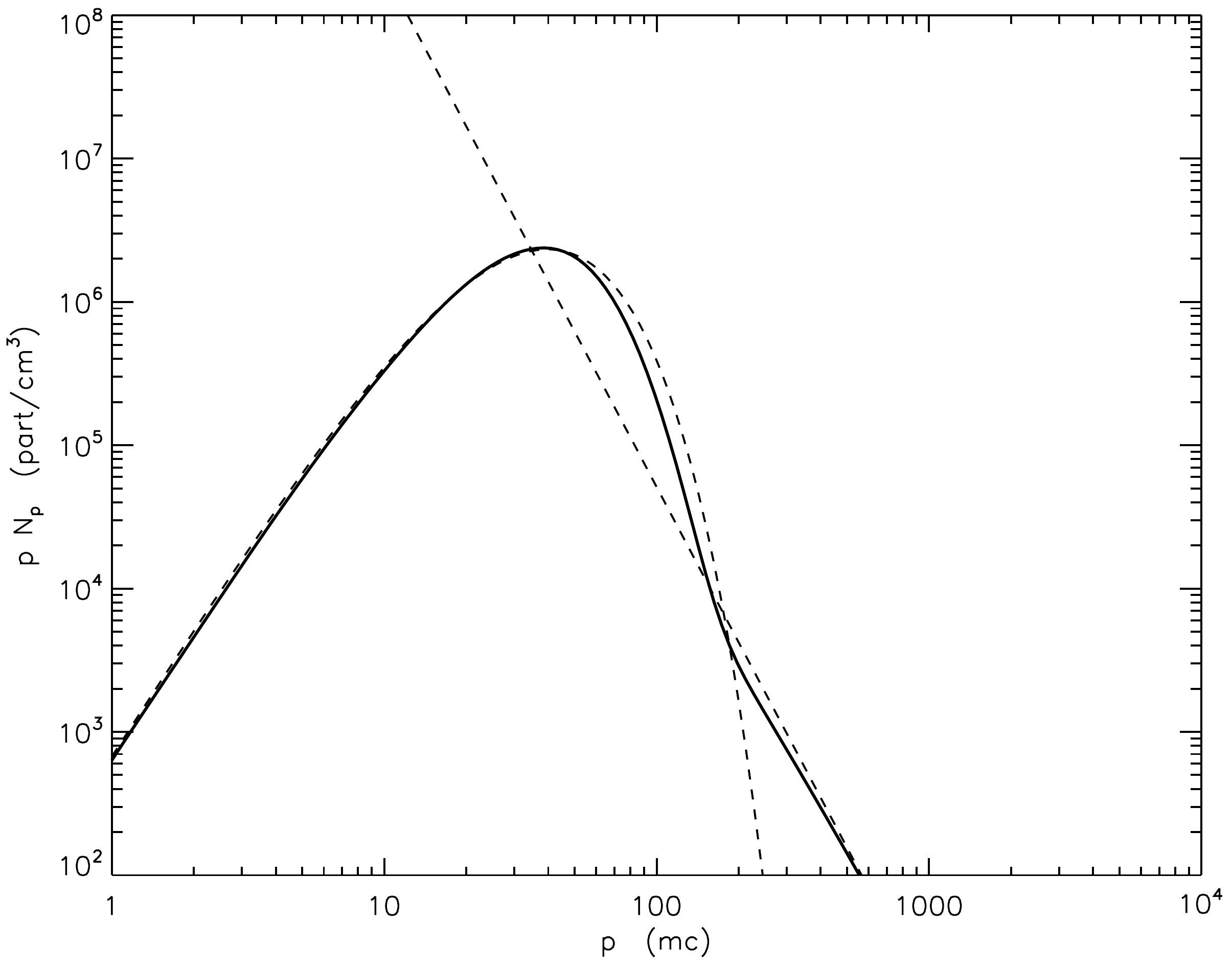}
 \caption{\textit{Quiescent spectrum from Sgr A* (top panel) and the associated lepton distribution (bottom panel) with thermal injection and escape (open configuration). The data points are the same as in the previous quiescent spectrum on Figure \ref{Quiescent00}}} 
\label{Quiescent33}
\end{figure}

For comparison within the open configuration, we want to reproduce the quiescent spectrum with the assumption that particles flows in and out of the emitting region. Figure \ref{Quiescent33} shows such a spectrum, which is similar to the simulated spectrum in Figure \ref{Quiescent00}. This spectrum is a realistic and acceptable solution for the quiescent state of Sgr A*. The magnetic field magnitude is 175 Gauss with a resulting plasma density of $3.3\times 10^{6} cm^{-1}$ that is also magnetically dominated with $\epsilon _k/\epsilon _b=0.08$.  We note that in this case, the thermal component does not really differ from the pure Maxwellian distribution (see bottom panel of Figure \ref{Quiescent33}). Indeed, in this range, radiative cooling is negligible, so that the balance between the thermal particle injection and the uniform particle escape produces a steady state distribution that is almost the pure Maxwellian. The total luminosity of this spectrum is L = $1.0 \times 10^{36}$ erg $\rm s^{-1}$, also equivalent to the previous quiescent fit. From this quiescent spectrum we next investigate the changes necessary in order to move to the flaring spectrum.

\begin{figure}
 \includegraphics[scale=0.35]{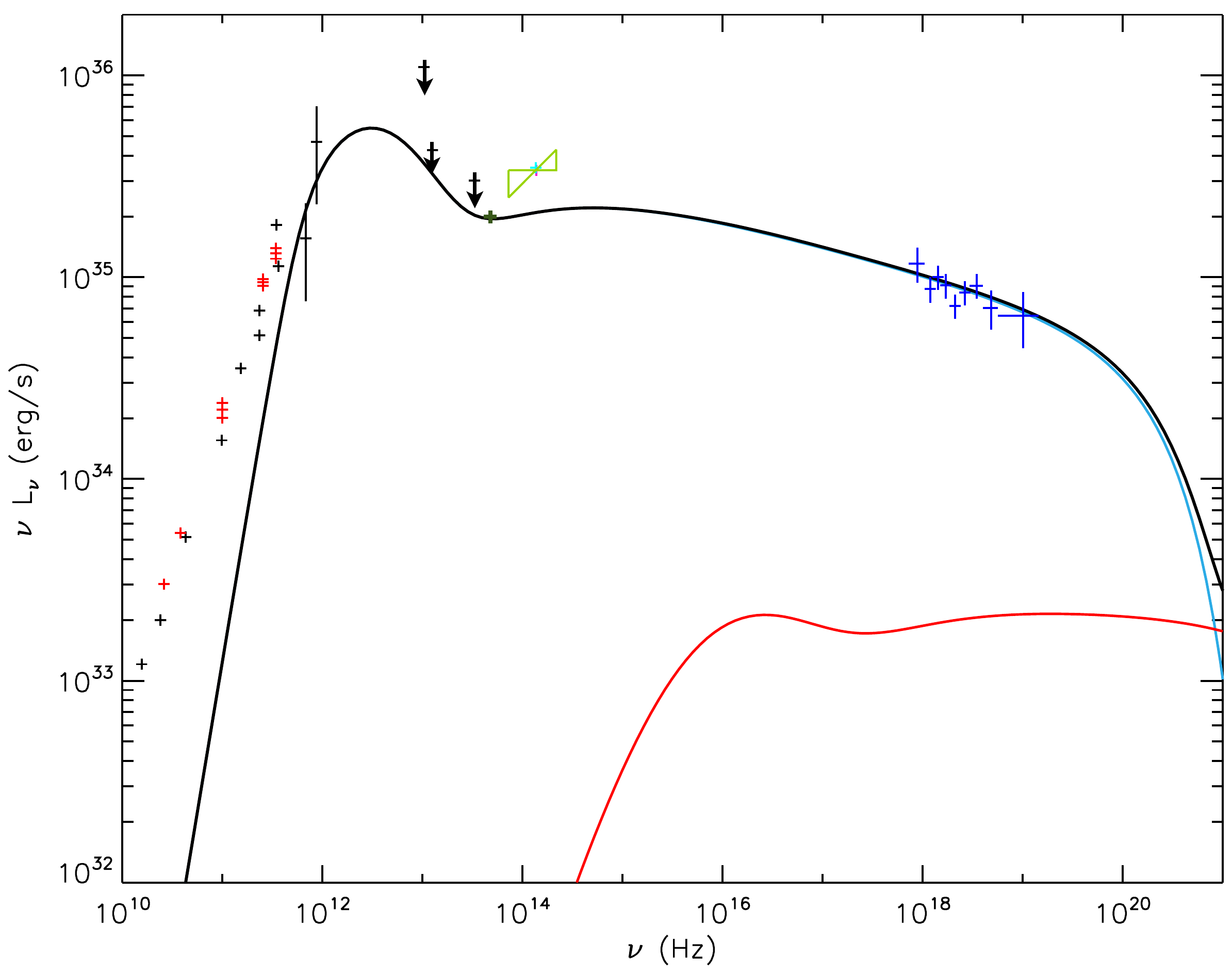}
 \includegraphics[scale=0.35]{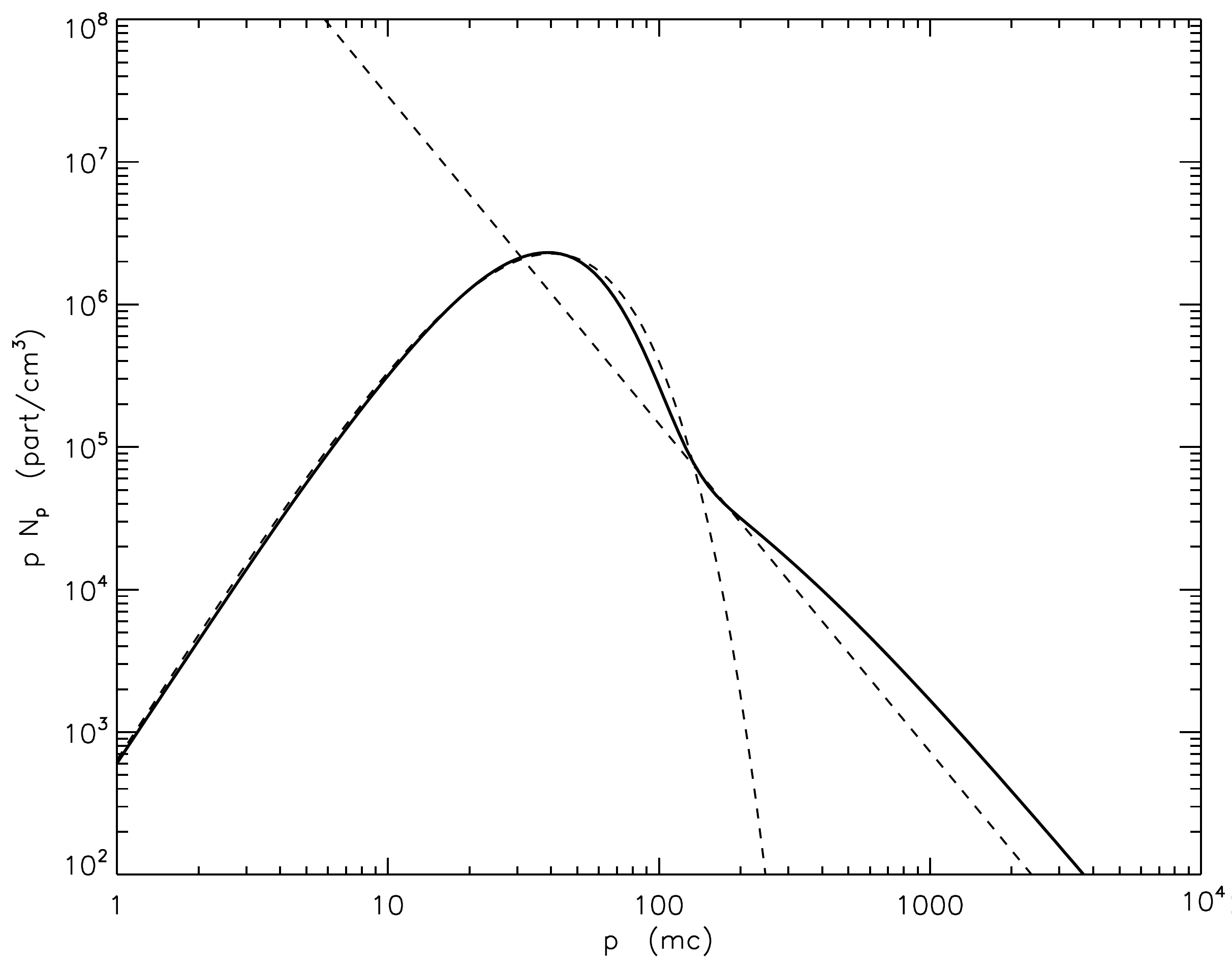}
 \caption{\textit{Flare spectrum from Sgr A* (top panel) and the associated lepton distribution (bottom panel) in the open configuration. The data points are the same as in the previous spectrum on Figure \ref{Flare00c}. The electron distribution shows also in dotted line, the pure Maxwellian and power-law curves as a comparison. }} 
\label{Flare33s_J21}
\end{figure} 

\begin{figure}
 \includegraphics[scale=0.35]{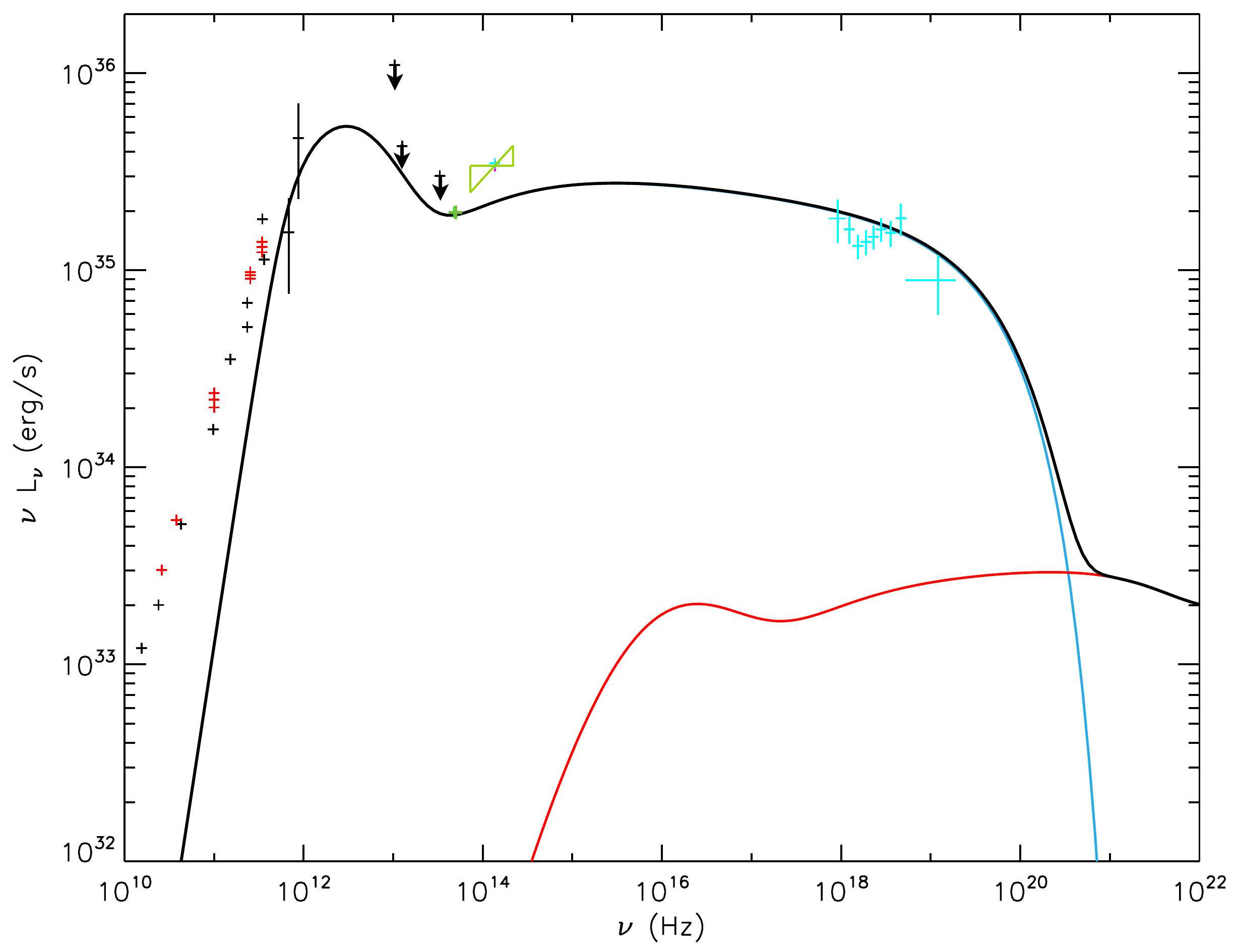}
 \includegraphics[scale=0.35]{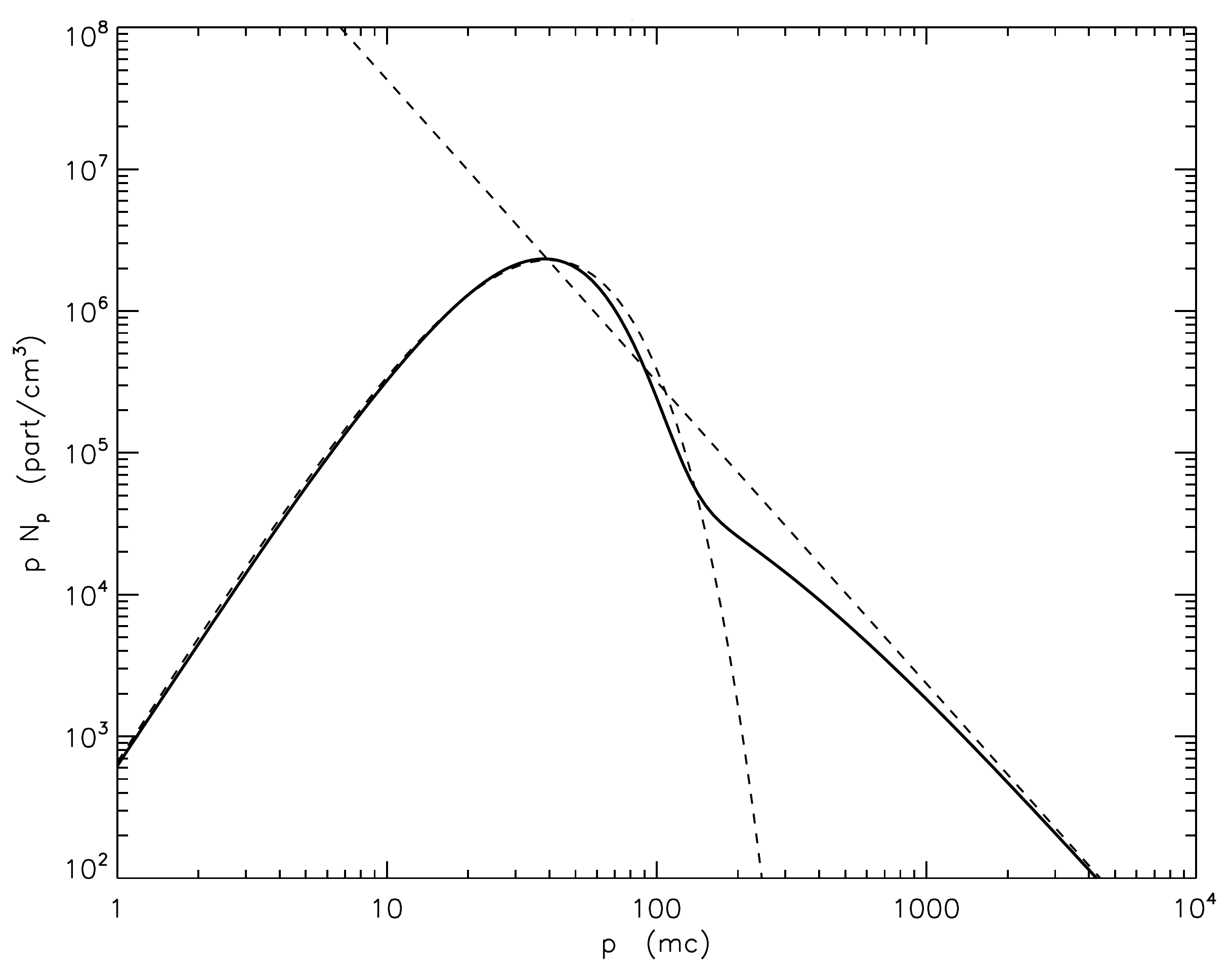}
 \caption{\textit{Flare spectrum from Sgr A* (top panel) and the associated lepton distribution (bottom panel) with thermal injection and escape. The data points are the same as in the previous flare spectra on Figures \ref{Flare00c}, \ref{Flare00s}, and \ref{Flare33s_J21}. The calculated electron distribution (full line) and the theoretical one (dotted line) as a comparison.}} 
\label{Flare33s_O17}
\end{figure}

Figure \ref{Flare33s_J21} shows a spectrum for a flaring state of Sgr A*, together with the lepton distribution. The flaring spectrum is dominated by non-thermal synchrotron and reproduce the NuSTAR July flare as well as an IR flare with a slope closer to the usually observed one (flat to slightly rising in the power spectrum). As expected, cooling breaks are observed in the lepton distribution and in the photon spectrum. These are not sharp but span at least one order of magnitude in frequency. The emitting region is the same as in the quiescent state and the magnetic field stays the same as well. The amount of injected particles is the same as in quiescence leading to a constant density. The only change in order to move from the quiescent to the flare spectrum is on the non-thermal component: the heating parameter $\rm l_{nth}$ is increasing by almost one order of magnitude, and the slope becomes flatter (from 3.6 to 2.3 during the flare) meaning that we have more particles in the higher energy part of the electron distribution. So, we must have some physical processes that accelerates the particles more efficiently in the flaring state and creates a harder non-thermal distribution. As a consequence the total luminosity increases, reaching $3.8 \times 10^{36} \rm erg \ s^{-1}$.

We can do the same exercise to reproduce the October NuSTAR flare. This is shown in Figure \ref{Flare33s_O17} for our best case scenario that is really similar to the model on Figure \ref{Flare33s_J21} with non thermal synchrotron with a cooling break responsible for the flare emission. This is not surprising as we explained earlier in the Data section 3.1, that both flares are not significantly different and should be modelled with a power-law shape as a fit. As for the July flare, the trigger of the event is on the non-thermal component of the lepton distribution that increases by a bit more than an order of magnitude with a prescribed slope of 2.1 which is a bit flatter than for the July flare. Beside that, the size of the emitting region,  the magnetic field, and the density are the same as for the other flare and the same as in quiescence. For the October flare model we have a slightly higher non-thermal power with a slightly flatter prescription for the acceleration, the total luminosity of this flare spectrum is $4.8 \times 10^{36} \rm erg \ s^{-1}$

\begin{figure}
 \includegraphics[scale=0.35]{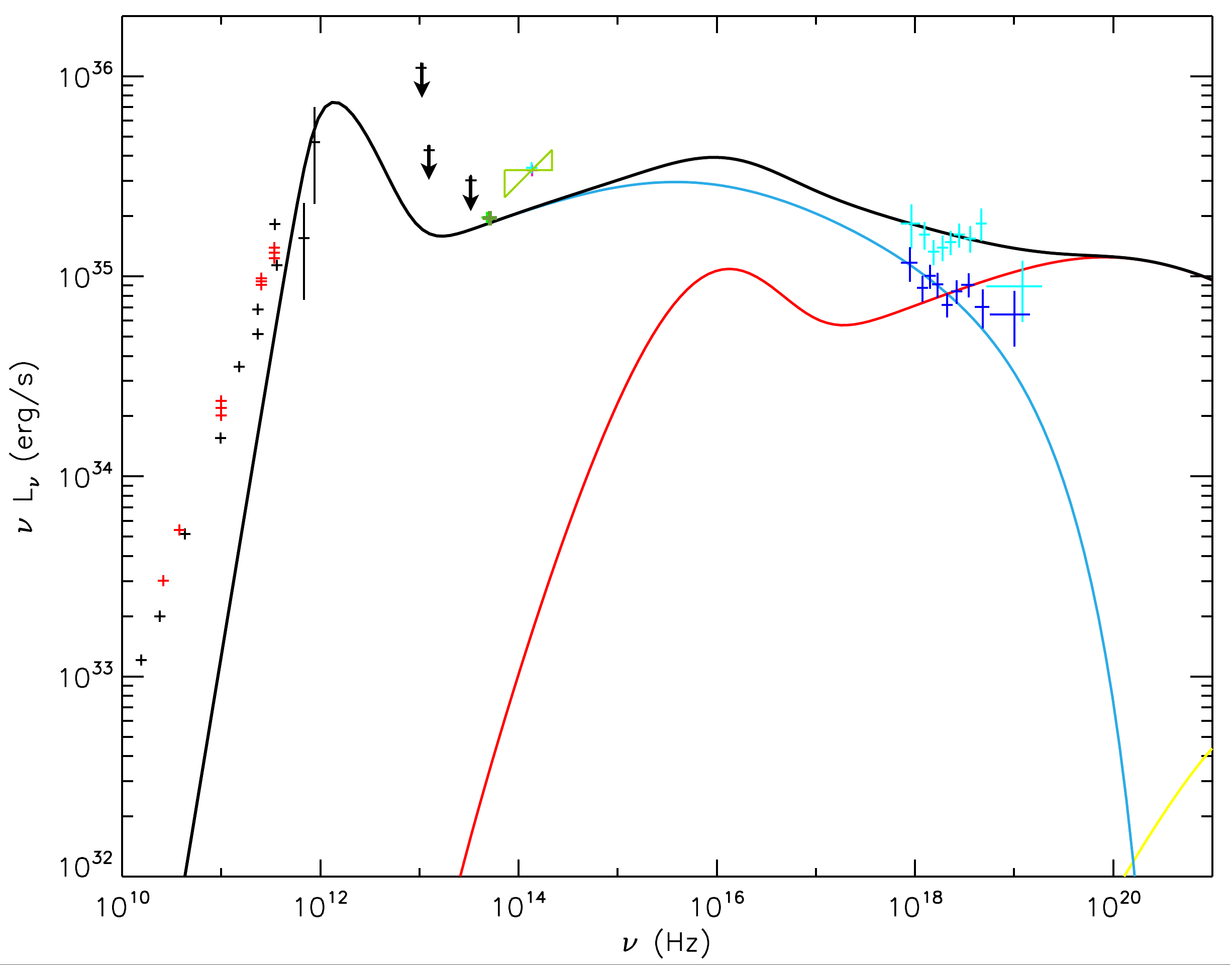}
 \includegraphics[scale=0.35]{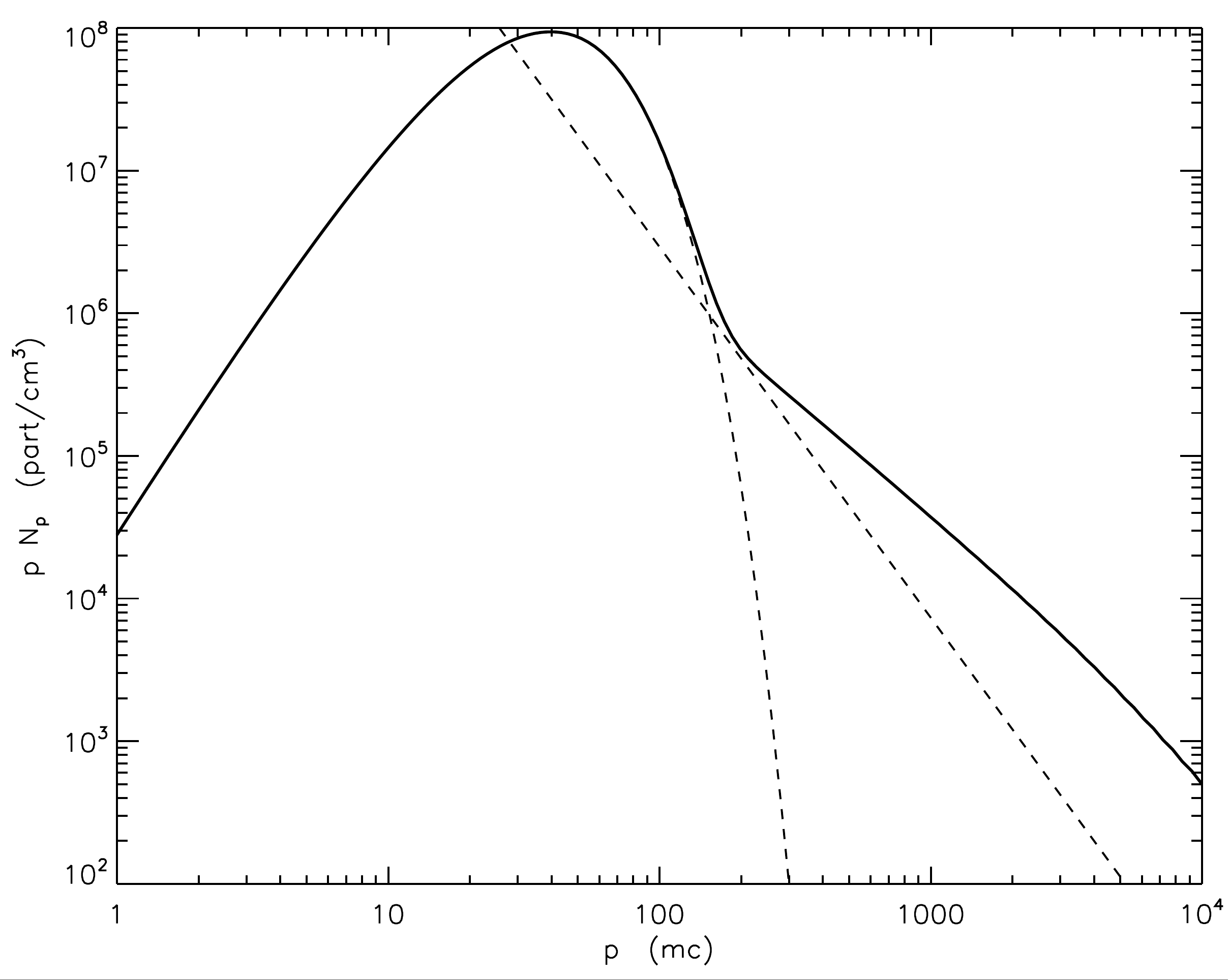}
 \caption{\textit{Flare spectrum from Sgr A* (top panel) and the associated lepton distribution (bottom panel) with thermal injection and escape. The data points are the same as in the previous flare spectra on Figures \ref{Flare00c}, \ref{Flare00s}, \ref{Flare33s_J21}, and \ref{Flare33s_O17}. The calculated electron distribution (full line) and the theoretical one (dotted line) as a comparison.}} 
\label{Flare33c}
\end{figure} 

We investigated an alternative scenario where the X-ray flares would be produced by synchrotron self Compton (SSC) emission, however we found that models that account for pure SSC as an emission mechanism have physical parameters that are hardly compatible with what we know of the central region density. Moreover it leads to a more complex scenario where the large scale magnetic field and the density of the medium need to be modified during the flare event. Nevertheless, inverse Compton emission could still be a non-negligible component of the overall spectrum, especially at high energies, assuming a weaker magnetic field during the flare and a higher density medium. Figure \ref{Flare33c} gives an illustration of some ``power-law'' shape X-ray emission that would be a combination of synchrotron and SSC. In this case the density has increased from $3 \times 10^{6}$ to $1 \times 10^{8}$ $\rm cm^{-1}$ and the magnetic field has dropped from 175 to 35 Gauss moving to a kinetic dominated flow with $\epsilon_k/\epsilon_b=97$.

%Figure \ref{Flare33s} shows a second possible model for the flare data from Sgr A*, together with the lepton distribution. The difference is that this spectrum is dominated by synchrotron emission, even though Compton emission is also significant. In this case, the only change in order to move from the quiescent to the flare spectrum is on the non-thermal component: the heating parameter $\rm l_{nth}$ is increasing by a factor six, and the slope becomes flatter (from 3.6 to 2.6 during the flare) meaning that we have more particles in the higher energy part of the electron distribution. So, we must have some physical processes that accelerates the particles more efficiently in the flaring state and creates a harder non-thermal distribution. As a consequence the total luminosity increases, reaching $4.1 \times 10^{36} \rm erg \ s^{-1}$.

%\subsection{Discussion}
We think that the best model for the flaring state of Sgr A* is the one produced by non-thermal synchrotron with a cooling break as seen on Figure \ref{Flare33s_J21} and \ref{Flare33s_O17} because the trends of the  multi-wavelength data are reproduced and only very few parameters need to be adjusted in order to move from the quiescent to the flaring state. This is especially true if we consider that the green ``bowtie'' is a typical IR slope. The non-thermal synchrotron emission is a simple and elegant solution of the flaring event observed by Chandra and NuSTAR because the overall state of the medium does not change dramatically (for instance the density and magnetic field is kept constant). The acceleration of the electrons leading to the more important and flatter non-thermal lepton distribution is the only modification, and this could be triggered by some plasma instabilities that are not modelled in details here. 

Even-though magnetic reconnections could also be the initial trigger, it can happen on very small scales and does not necessary lead to a drop of the global magnetic field magnitude. A possible sudden increase of the density (as in model \ref{Flare33c}) can be interpreted as an accretion rate fluctuation, however such fluctuations of more than an order of magnitude are most likely not happening every day in the Galactic Center and would be difficult to interpret.  The NuSTAR data being consistent with a power-law shape to higher energies points also in favour of the synchrotron scenarios as in models \ref{Flare33s_J21} and \ref{Flare33s_O17}.  

The sub-millimeter part of the spectrum is really stable: comparing the quiescent state on Figure \ref{Quiescent33} with the sub-millimeter part of the spectrum on Figure \ref{Flare33s_J21} we have exactly the same contribution around $10^{12}$ Hz. This is mainly due to the fact that the magnetic field is kept constant and the injected population is also constant. This configuration gives us in return a constant density.  We have to keep in mind that the data are not simultaneous, nevertheless it is an interesting exercise trying to model several wavelength observations in the same time. In the future simultaneous observations are going to be very important for this kind of multi-wavelength study. 

\begin{figure}
 \includegraphics[scale=0.26]{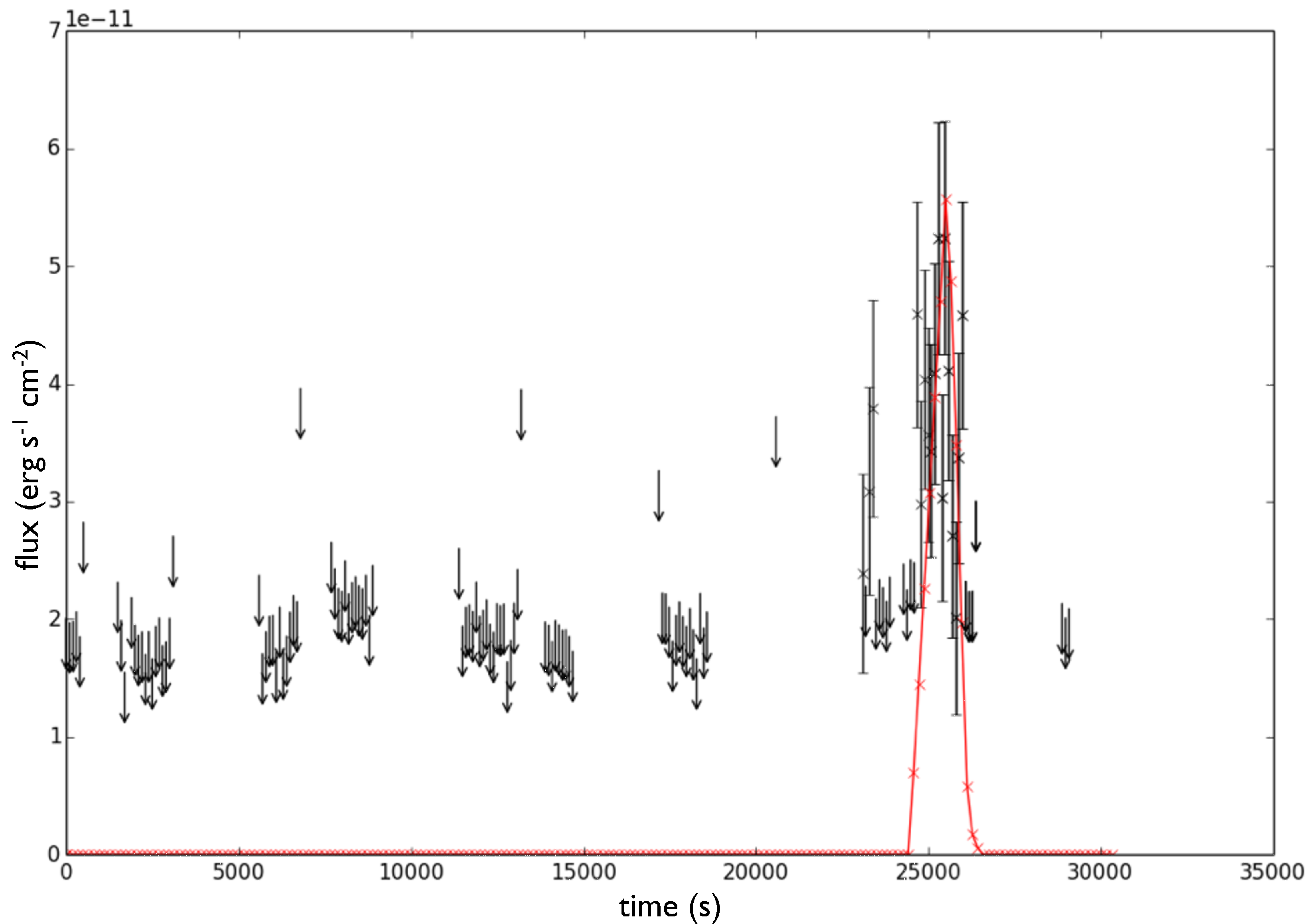}
 \caption{\textit{NuSTAR X-ray light curve of the flare event of the 21st of July 2012. The black data points is the unabsorbed flux between 3 and 79 keV. When the detection was not strong enough to be significant, we have only plotted upper limits of three sigma (arrows). See \citet{barriere14} for the observed light-curve. The red curve represents the X-ray light curve from our quiescent spectrum model in Figure \ref{Quiescent33} to the flare spectrum model in Figure \ref{Flare33s_J21}.}} 
\label{LightCurve}
\end{figure} 

We then looked at the time evolution between Figure \ref{Quiescent33} and \ref{Flare33s_J21} to reproduce the X-ray light curve of the July NuSTAR flare.  Using our self-consistent calculations, we can also model the time-dependent particle evolution in order to reproduce the flare light-curves. This approach has also been considered by \cite{doddseden10} in the one zone cell approximation but for a given power-law distribution. 
The cooling time-scales being very short compared to the flare duration, the time evolution is entirely governed by the physics of the acceleration processes that are not clearly defined. Figure \ref{LightCurve} shows the reproduction of the X-ray light curve between 3 and 79 keV for the same parameter setting as for Figure \ref{Quiescent33} and \ref{Flare33s_J21}.  During the flare, the non-thermal parameter $\rm l_{nth}$ evolves linearly with time from the quiescent value $10^{-5}$ to a maximum value, such that the averaged value over the flare duration is $9.8 \times 10^{-5}$ as in our flare spectrum \ref{Flare33s_J21}. It reaches a maximum value at the peak, and decreases back immediately with a linear dependence. The slope of the accelerated particles is set to 2.28 during the flare event as in our flare spectrum \ref{Flare33s_J21}, and to 3.60 in quiescence. We note that before the flare event, Sgr A* is not detected by the X-ray satellite NuSTAR because  it is too faint, and embedded in the diffuse/unresolved emission, in this case we simply plotted  3 $\sigma$ upper limits.  

%Compared to our model, it seems that the real flare has a higher flux, this is because we based our time dependent simulation on the average spectrum of Figure \ref{Flare33s_J21}. Namely, we took the physical value for the power of non-thermal acceleration $\rm l_{nth}$ from the flare spectrum as the averaged value over the light-curve. And for the prescribed slope $s$, we are reaching the value for the flare only at the peak. In reality, the harder spectrum must remain for a longer time during the flare leading to broader emission as seen with NuSTAR.  

\section{Conclusion and Outlook}

We are able to reproduce the quiescent spectrum of Sgr A* in two different scenarios: considering that the accretion process is very slow and that the same particles remain a long period of time in the emitting region, and considering that the accretion process is very efficient, with and accretion velocity close to the speed of light; so that particles only remain in the emitting region on short time scales comparable to the radiative time scales. To model the flaring state however, we favour the second scenario, that allows a better interpretation of the sub-millimetre and infra-red part of the spectrum. The flaring state spectrum is best reproduced by a plasma that has the same low magnetic field as in quiescent, and the same amount of injected particles. More efficient non-thermal heating processes are responsible for the flaring event, and a flatter non-thermal distribution of electrons is present. Besides this change, all other parameters stay the same when moving from the quiescent to the flaring spectrum (Figure \ref{Quiescent33} and \ref{Flare33s_J21}). 
Our conclusions are in good agreement with \citep{doddseden10} who also favoured non-thermal synchrotron processes and a cooling break in order to explain the observed IR and X-ray flares. However, in our study we do not make the hypothesis of magnetic reconnection as an energy power for the flares, and our conclusions do not favour this particular hypothesis. As in our best case scenario (Figures \ref{Quiescent33} and \ref{Flare33s_J21} or \ref{Flare33s_O17}), the magnetic field is not required to drop significantly. An important drop in the magnetic field amplitude has also important consequences on the sub-millimetre and thermal part of the spectrum that we also model here, other parameters have then to be carefully adjusted in order to maintain the sub-mm shape in reasonable values, so we think other acceleration mechanisms are more likely to be happening. Reconnection mechanisms could also occur in very localised regions, and particles would diffuse away from the reconnection sites and radiate in a field which has not reconnected, so we would not notice any significant global drop of the magnetic field amplitude.
\begin{figure}
 \includegraphics[scale=0.35]{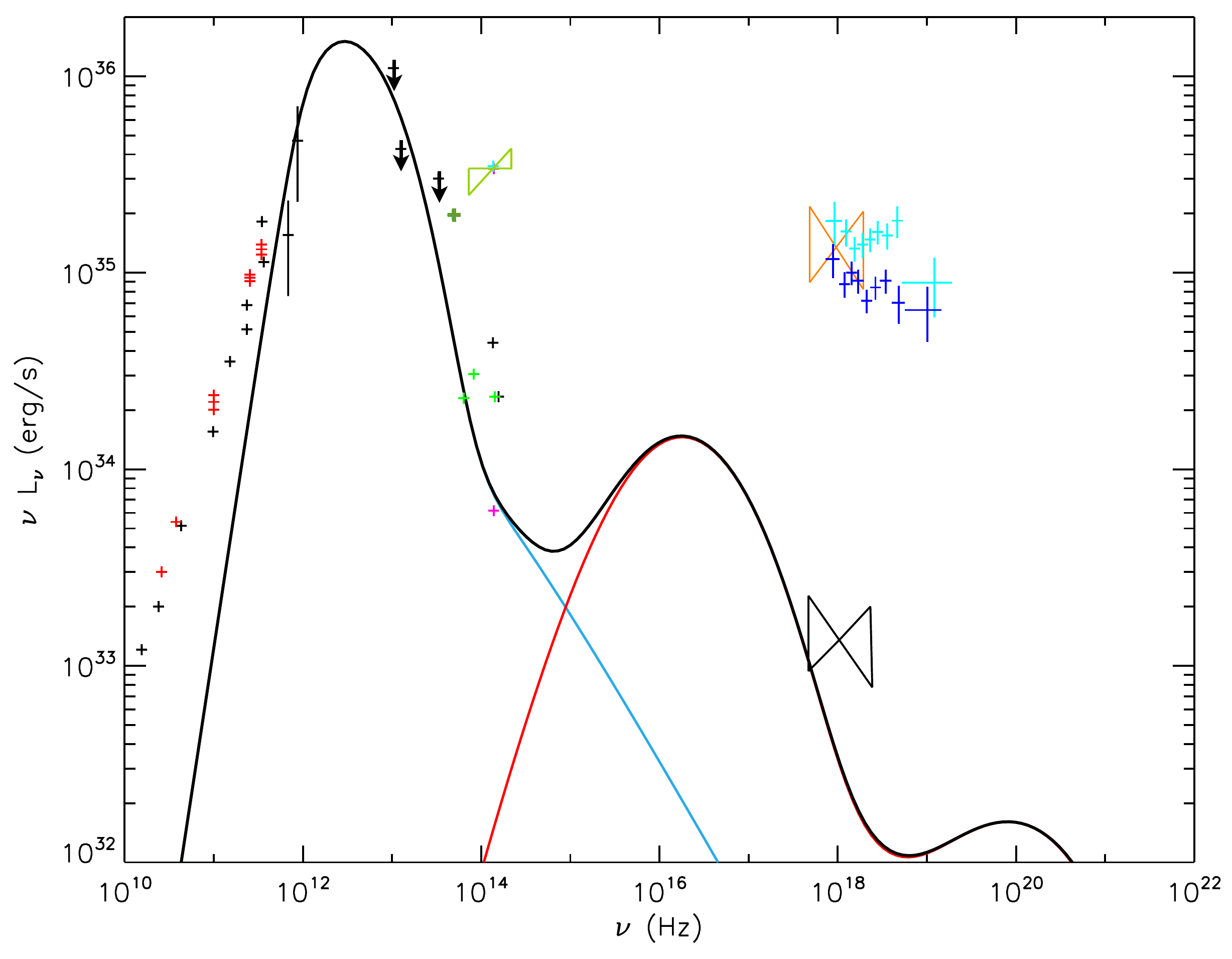}
 \includegraphics[scale=0.35]{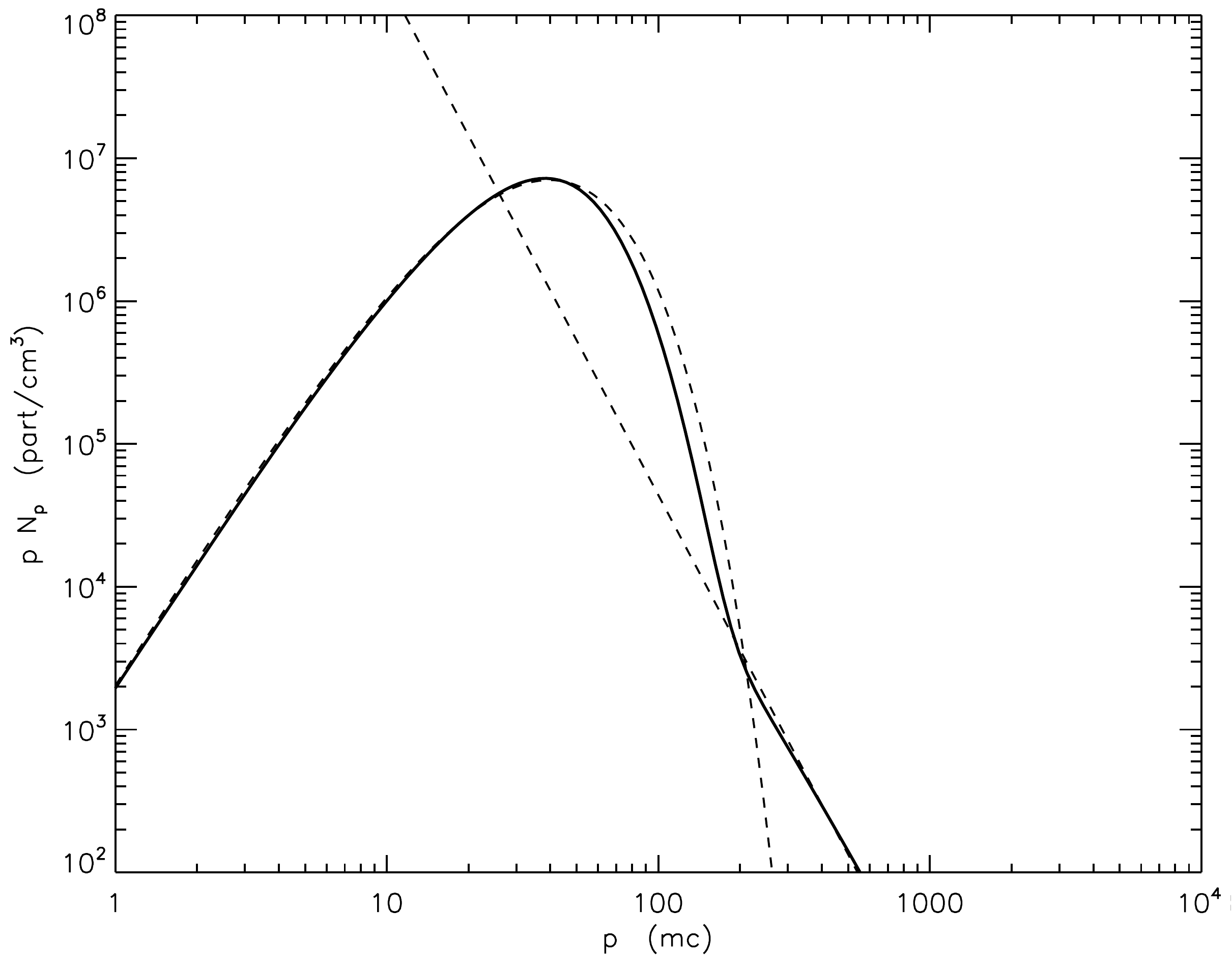}
 \caption{\textit{Quiescent spectrum from Sgr A* with the same conditions as in Figure \ref{Quiescent33} but assuming an increase of the density by a factor three. All the observed data (of quiescent and flaring state) has been kept on the figure. The associated lepton distribution is shown on the bottom panel.}} 
\label{Quiescent33_x3}
\end{figure} 
In our study, we end up with a plasma density of $3.3\times 10^{6}$ particles per cubic centimetre, which is a reasonable value according to observations and theoretical work. But what would happen to the quiescent spectrum (Figure \ref{Quiescent00} or \ref{Quiescent33}) if the density increases by a factor three as expected to happen now when the cloud G2 is falling into the Galactic Center? As reported by \citet{gillessen12}, a dense gas cloud approximately three times the mass of Earth is falling into the accretion zone of Sgr A*, but nothing noticeable has been observed yet from Sgr A*. Figure \ref{Quiescent33_x3} represents such a prediction, it has exactly the same settings as the model described on Figure \ref{Quiescent33} but the density is three time higher (we have more particle injection). The model predicts a flux increase in the sub-mm bump ($10^{12}$ -- $10^{13}$ Hz), however the current emission is not well constrained in this band. If the source stays in quiescence, we do not expect a particular increase in the IR, and we have some emission in the ultra-violet due to the first Compton component that is unfortunately not detectable. Even in the X-ray, if the increasing density by a factor three does not trigger a flare event, we do not expect a significant increase from the quiescent X-ray level. Overall it could well be that we are not detecting any striking changes.

%Note that we reach a density of $1.4\times 10^8$ particles/$\rm cm^3$, which is the same as the density obtained in model \ref{Flare33c}, representing the flaring state with a Compton emission in the X-ray. It could be difficult to disentangle an increase of the density in the quiescent state with a flaring episode, as the two could be linked. Indeed, to move from the quiescent spectrum on Figure \ref{Quiescent33} to the flaring spectrum on Figure \ref{Flare33c} we have a density jump, but such a change is not necessary to reproduce a flare. The model used to generate the spectrum on Figure \ref{Flare33s} that reproduces best the data in the flaring state, has the same density as in the quiescent state. Moreover, a simple increase of the density does not produce an IR flare. Another mechanism must be powering the flare events by accelerating the particles to higher energies with a non-thermal distribution.

\section*{Acknowledgments}

We acknowledge support from The European Community’s Seventh Framework Programme (FP7/2007-2013) under grant agreement number ITN 215212 Black Hole Universe.
We also acknowledge support from the ``Nederlandse Onderzoekschool Voor Astronomie'' NOVA Network-3 under NOVA budget number R.2320.0086.\\
SM and SD gratefully acknowledge support from a Netherlands Organization for Scientific Research (NWO) Vidi Fellowship.\\
RB and JM acknowledge financial support from both the french National Research Agency  (CHAOS project ANR-12-BS05-0009) and the Programme National Hautes Energies.
%We acknowledge support from The European Community's Seventh Framework
%Programme (FP7/2007-2013) under grant agreement number ITN 215212
%Black Hole Universe.  SM and SDi gratefully acknowledge support from a
%Netherlands Organization for Scientific Research (NWO) Vidi
%Fellowship.   This work was also partially supported by the National Science
%Foundation under grants AST 0807385 and PHY11-25915 and through TeraGrid resources
%provided by the Texas Advanced Computing Center (TACC).  We thank SARA
%Computing and Networking Services (www.sara.nl) for their support in
%using the Lisa Computational Cluster.  PCF acknowledges support of a
%High-Performance Computing grant from Oak Ridge Associated
%Universities/Oak Ridge National Laboratory.

%\bibliographystyle{mn2e}
%\bibliography{refs}
%\bsp

\label{lastpage}
\end{document}